\documentclass[12pt]{article}
\usepackage{amsbsy}
\usepackage{amsfonts}
\usepackage{amsmath}
\usepackage{amssymb}
\usepackage{apacite}
\usepackage{setspace}
\usepackage{graphicx}
\usepackage{bm,multicol}
\usepackage[english]{babel}
\usepackage{natbib}
\usepackage[T1]{fontenc}
\usepackage[utf8]{inputenc}
\usepackage{authblk}
\usepackage{xcolor}
\usepackage{geometry}
\newcommand{\newc}{\newcommand}
\newc{\N}{\mbox{N}}
\newc{\1}{\bf{1}}

\begin{document}
\title{Bayesian Testing of Scientific Expectations Under Exponential Random Graph Models}

\author{J. Mulder, N. Friel, \& P. Leifeld}
\date{}

\maketitle

\begin{abstract}
\noindent The exponential random graph (ERGM) model is a commonly used statistical framework for studying the determinants of tie formations from social network data. To test scientific theories under the ERGM framework, statistical inferential techniques are generally used based on traditional significance testing using $p$-values. This methodology has certain limitations, however, such as its inconsistent behavior when the null hypothesis is true, its inability to quantify evidence in favor of a null hypothesis, and its inability to test multiple hypotheses with competing equality and/or order constraints on the parameters of interest in a direct manner. To tackle these shortcomings, this paper presents Bayes factors and posterior probabilities for testing scientific expectations under a Bayesian framework. The methodology is implemented in the R package \texttt{BFpack}. The applicability of the methodology is illustrated using empirical collaboration networks and policy networks.
\end{abstract}

\noindent \textbf{Keywords:} Bayesian hypothesis testing, Bayes factors, exponential random graph models, $g$-priors.

\section{Introduction}
The exponential random graph model (ERGM) is one of the most widely used statistical frameworks for explaining and predicting the formation of ties between actors in a network using the characteristics of the actors and the endogenous characteristics of the network. The framework has resulted in many new insights in various scientific fields, including brain networks in neuroscience \citep{simpson2011exponential}, board interlocks between firms in management research \citep{kim2016understanding}, genetic and metabolic networks in biology \citep{saul2007exploring}, and adolescent friendship networks in sociology \citep{goodreau2009birds}.

A statistical analysis under an ERGM is typically executed by first obtaining the maximum likelihood estimates of the network parameters, which quantify the relative importance of the predictor variables, and the corresponding standard errors, which quantify the statistical uncertainty of the estimates given the (limited) information in the network. Subsequently, to test which predictor variables have a statistical effect on the formation of ties in the network, classical $p$-values are often used. When testing a null hypothesis of whether an ERGM parameter equals zero or not, the $p$-value is defined as the probability of observing effects that are at least as extreme as the estimated effect from the observed network under the assumption that the true effect equals zero. Despite the ubiquity of this approach in scientific practice however, there is an increasing debate about the usefulness of $p$-values for statistical hypothesis testing \citep{benjamin2018redefine,mulder2016editors,Rouder:2009}. For applied researchers, a potential issue may be its difficult interpretation as $p$-values are often misinterpreted as error probabilities \citep{greenland2016statistical}. Besides this practical issue, the $p$-value also has more fundamental limitations, which also restrict its usefulness for applied social network research under the ERGM framework.

A first limitation is that $p$-values cannot be be used to quantify the evidence in the data in favor of a null hypothesis, but it can only falsify a null hypothesis. This is a limitation when the null hypothesis reflects an important scientific expectation or theory. \cite{leifeld2012information}, for instance, investigated whether political actors with similar policy preferences form information exchange ties in policy networks. They hypothesized that preference similarity had no additional effect on the formation of ties when covariates based on institutional, relational, and social opportunity structures were also present in the model. When they tested this hypothesis, a $p$-value larger than the significance level was obtained. The difficulty with this non-significant outcome is that on the one hand the authors could not claim that there is enough evidence in the data to reject the null, and on the other hand it could not be stated that there is evidence in favor of the null (because using a $p$-value we can only falsify a null). A cause of this problem is that the $p$-value cannot distinguish between the absence of evidence in the data and evidence of absence of an effect \citep{dienes2014using}. Therefore, non-significant results cannot guide us about conclusions regarding scientific theories.

A second limitation is that the $p$-value is statistically inconsistent when the null is true. This is a direct consequence of the central property that the $p$-value is uniformly distributed under the null. This property is crucial in order to control the type I error probability of incorrectly rejecting the null using a prespecified significance level. A consequence is, however, that there is always a fixed probability (equal to the significance level) of incorrectly rejecting a true null hypothesis, even when the size of the sampled network goes to infinity. Practically, this implies that when collecting information from a huge network (which may be very costly and time consuming) and when there is no effect of a key predictor variable on the formation of ties, there is still a strictly positive chance to draw an incorrect conclusion, which is typically equal to .05.


A final limitation we mention here is that $p$-values are not designed for testing multiple related hypotheses. Thus, when multiple hypotheses are formulated based on competing scientific expectations, it is not straightforward to test these hypotheses in a direct manner against each other using $p$-values. Moreover, when a hypothesis also contains order constraints on the parameters of interest, which is often the case in applied research as expectations are often formulated using `larger than' or `smaller than' statements \citep{Hoijtink:2011}, $p$-values can only be used for a limited set of hypothesis tests \citep{Silvapulle:2004}. For example, $p$-values are not available for directly testing nonnested order-constrained hypotheses, e.\,g., $H_1:\beta_1\le\beta_2\le\beta_3$ versus $H_2:\beta_1>\beta_2>\beta_3$. Testing order-constrained hypotheses using post-hoc tests is also problematic for another reason: We can obtain conflicting conclusions (e.\,g., `$\beta_1=\beta_2$' and `$\beta_1=\beta_3$' cannot be rejected but `$\beta_2=\beta_3$' can be rejected), which is problematic for applied statistical practice. Nevertheless, testing order-constrained hypotheses is very relevant for social network research. For example, \cite{leifeld2018polarization} investigated the effect of collegial relationships on the tendency to co-author papers in scientific collaboration networks. Supervisor--supervisee pairs were expected to show a stronger tendency to collaborate on publications than colleagues only working in the same team. In turn, same-team colleagues were expected to exhibit a stronger tendency to collaborate than colleagues merely working in the same institution. And finally, people in the same institution were expected to co-author at higher rates than people affiliated with different institutions. In such cases, inferential techniques for order-constrained hypotheses are desirable.

To resolve these issues and limitations of $p$-values, this paper proposes a flexible and fast Bayesian test under the ERGM framework. The methodology uses Bayes factors and posterior probabilities, which (i) can be used to quantify the relative evidence in the data in favor of a null hypothesis, (ii) are consistent under general conditions, and (iii) are broadly applicable for testing multiple hypotheses simultaneously and for testing hypotheses with order constraints on the parameters of interest. A weakly informative `unit-information prior' is proposed which builds on the well-known $g$-prior approach of \cite{Zellner:1986}. This prior can be used in an automatic fashion for Bayesian hypothesis testing and estimation without requiring external prior information. Note that prior specification under Bayesian ERGMs is still a largely unexplored topic in the ERGM literature (a notable exception is the conjugate prior of \cite{yin2022highly} for Bayesian estimation). The computation of the resulting Bayes factors is extremely fast due to its reliance on Gaussian approximations of the posterior (following large sample theory). Furthermore, Bayes factors can be computed for any number of hypotheses using only one run of the `\texttt{bergm}' function from the \texttt{Bergm} package \citep{caimo2014bergm}. To facilitate its usability, the methodology is implemented in the R package \texttt{BFpack} \citep{mulder2021bfpack}. The main function \texttt{BF} only requires a fitted (classical or Bayesian) ERGM using either the \texttt{ergm} package \citep{hunter2008ergm,krivitsky2023ergm} or the \texttt{Bergm} package \citep{caimo2014bergm}.

In summary, the current paper makes three important contributions to the literature on Bayesian ERGMs. First, a novel prior specification method is proposed which can be used in an automatic fashion. Second, a fast method is proposed for computing Bayes factors between many hypotheses. Third, a broad class of multiple hypothesis tests can be executed using equality and/or order constraints on the ERGM coefficients.

The paper is organized as follows. Section 2 provides some background information about ERGMs and about the formulation of hypotheses which may involve order constraints on the parameters of interest. Two motivating empirical applications are then discussed in the context of scientific collaboration networks and information exchange in policy networks. Section 3 describes the methodology for Bayesian hypothesis testing under ERGMs and the implementation in the R package \texttt{BFpack} \citep{mulder2021bfpack}. In Section 4, the numerical behavior of the test is illustrated when testing a null hypothesis (and compared with classical $p$-values) and when testing order-constrained hypotheses. In Section 5, we apply the methodology in two empirical applications. Section 6 ends the paper with concluding remarks and a discussion.

\section{Statistical hypotheses under the ERGM framework} \label{section_ergm}

Under an ERGM, the probability of an adjacency matrix $\textbf{Y}$ of a network of $N$ actors, where $Y_{ij}=1$ denotes a tie (or edge) between actors $i$ and $j$, and $Y_{ij}=0$ denotes the absence of a tie between $i$ and $j$, is given by
\begin{equation}
p(\textbf{Y}|\bm\beta,\textbf{X}) = \frac{\exp\{\bm\beta^{\top}\textbf{s}(\textbf{Y},\textbf{X})\}}
{\sum_{\textbf{Y}'\in\mathcal{Y}}\exp\{\bm\beta^{\top}\textbf{s}(\textbf{Y}',\textbf{X})\}},
\end{equation}
where $\mathcal{Y}$ denotes the set of all possible adjacency matrices that can be observed and $\textbf{s}(\textbf{Y},\textbf{X})$ denotes a vector of $K$ sufficient statistics which are assumed to explain the connectivity in the network given the endogenous characteristics of the network $\textbf{Y}$. The sufficient statistics within $\textbf{s}$ could include the tendency to form ties or the tendency to form triangles as well as the characteristics of the actors summarized in $\textbf{X}$. 
The coefficients in the vector $\bm\beta$ of length $K$ quantify the relative importance of the $K$ sufficient statistics in the formation of ties in the network.

An insightful way to formulate an ERGM is by considering the conditional probabilities of forming a tie for dyad $(i,j)$ given the rest of the graph \citep{wasserman1996logit}. The ERGM then comes down to a linear model for the log odds that a tie is present:
\begin{equation}
\log\left( \frac{\text{Pr}(Y_{(ij)}=1|\textbf{Y}_{-(ij)},\bm\beta,\textbf{X})}{\text{Pr}(Y_{(ij)}=0|\textbf{Y}_{-(ij)},\bm\beta,\textbf{X})}\right)
= \log\left( \frac{\text{Pr}(\textbf{Y}^{+}_{-(ij)}|\bm\beta,\textbf{X})}{\text{Pr}(\textbf{Y}^{-}_{-(ij)})|\bm\beta,\textbf{X}}\right)
= \bm\beta^{\top}\bm\delta_{(ij)}(\textbf{Y},\textbf{X})
\label{condlike}
\end{equation}
where $\textbf{Y}_{-(ij)}$ denotes the graph $\textbf{Y}$ where the tie of dyad $(i,j)$ is omitted, and $\textbf{Y}^{+}_{-(ij)}$ and $\textbf{Y}^{-}_{-(ij)}$ denote the graph where a tie is present and absent for dyad $(i,j)$, respectively, and $\bm\delta_{(ij)(\textbf{Y},\textbf{X})}$ is the vector of change statistics depending on whether a tie is present or absent for dyad $(i,j)$, i.e.,
\begin{equation}
\bm\delta_{(ij)}(\textbf{Y},\textbf{X})=
\textbf{s}(\textbf{Y}^{+}_{-(ij)},\textbf{X})-\textbf{s}(\textbf{Y}^{-}_{-(ij)},\textbf{X}).
\label{changestats}
\end{equation}

In practice, applied network researchers are often interested in testing the relative importance of the sufficient statistics to form ties in the network. Such hypotheses can be translated to competing hypotheses with equality and/or order constraints on the coefficients of the following form:
\begin{equation}
H_t:\textbf{R}_t^E\bm\beta=\textbf{0} ~ \& ~ \textbf{R}_t^O\bm\beta>\textbf{0},
\label{Ht}
\end{equation}
for $t=1,\ldots,T$, where the rows of the matrix for the equality constraints, $\textbf{R}_t^E$, and rows of the matrix of the one-sided (or order) constraints, $\textbf{R}_t^O$, are permutations of $(1,0,\ldots,0)$, $(-1,0,\ldots,0)$, or $(1,-1,0,\ldots,0)$. For example, in the case of a vector of 4 ERGM parameters, i.e., $\bm\beta^{\top}=(\beta_1,\beta_2,\beta_3,\beta_4)$, the matrix multiplication of $(0,1,0,0)\bm\beta>0$ would correspond to the one-sided constraint $\beta_2<0$, and the matrix multiplication of $(0,-1,0,1)\bm\beta>0$ would correspond to $\beta_4>\beta_2$. If a hypothesis only contains equality constraints, the matrix $\textbf{R}_t^O$ is not specified. The same holds if only order constraints are formulated, and then $\textbf{R}_t^E$ does not need to be formulated. Below, we provide more examples of these matrices for two empirical social network applications.

\subsection*{Application A: Information exchange in policy networks}
The first motivating application is the question whether political actors who share similar policy beliefs and political preferences also exchange information in policy networks. \citet{leifeld2012information} employed an ERGM to examine the institutional, relational, and social opportunity structures guiding the exchange of information among political actors in a national-level policy network on a contested issue. A policy network is comprised of the interest groups, government agencies, scientific bodies, and other organizations exerting influence over the policy formulation and decision-making process.
Actors can exchange technical and scientific or political and strategic information. Explaining under what conditions actors collaborate is instrumental for understanding why certain bills are adopted and why they contain specific elements, for example leaning towards interest group positions.

\citet{leifeld2012information} found that any two actors' similarity in policy preferences explained political/strategic information exchange between them. But this effect vanished when opportunity structures like shared committee memberships or shared partners were included in the model. This was seemingly at odds with the prior literature, which had posited that collaboration takes place at higher rates within ideologically homogenous regions of the network, not across political divides \citep[e.\,g.,][]{henry2011belief}. Would the lack of statistical significance once control variables were included have been evidence \emph{against} an association between preference similarity and tie formation? \citet{leifeld2012information} concluded that the effect of preference similarity was ``absorbed'' by opportunity structures, meaning that opportunity structures were mainly conducive to tie formation between actors with similar preferences, hence preferences per se were no longer a significant predictor themselves. The original study did not have access to a hypothesis testing framework that would have permitted to quantify evidence in favor of the hypothesis that no association existed, controlling for other effects. Below in Section~\ref{section_app}, we review this question using the original specification and test the hypothesis that there was no effect of preference similarity on information exchange in the presence of control variables.

\paragraph{$\textbf{H}_1$: No effect of preference similarity.}
The first hypothesis assumes that preference similarity has no effect on political/strategic information exchange in a policy network while controlling for the number of edges, governmental target, scientific source, number of common committees, scientific communication, interest group homophily, influence attribution, geometrically weighted edge-wise shared partners (GWESP with a decay parameter of 0.1), geometrically weighted dyad-wise shared partners (GWDSP with a decay parameter of 0.1), and reciprocity. Mathematically, the hypothesis can be written as
\[
H_1:\beta_{\text{pref sim}}=0
\]
Furthermore, the full vector is given by
\[
\begin{aligned}
\bm\beta = ( & \beta_{\text{edges}},\beta_{\text{pref sim}},\beta_{\text{gov target}},\beta_{\text{sci source}},\beta_{\text{committee}},\beta_{\text{sci comm}},\\
             & \beta_{\text{int group homophily}},\beta_{\text{infl attr}},\beta_{\text{GWESP}},\beta_{\text{GWDSP}},\beta_{\text{reciprocity}})^{\top}.
\end{aligned}
\]
Consequently, equality-constrained hypothesis $H_1:\beta_{\text{pref sim}}=0$ can be written using the matrix notation in \eqref{Ht} as $\textbf{R}_1^E=(0,1,0,\ldots,0)^{\top}$.

\paragraph{$\textbf{H}_2$: Non-zero effect of preference similarity (complement).}
The second hypothesis assumes that preference similarity has a non-zero effect on information exchange while controlling for the same covariates as under $H_1$. Mathematically, the hypothesis can be written as
\[
H_2:\beta_{\text{pref sim}}\not=0.
\]

Note that even though hypotheses $H_1$ and $H_2$ are often denoted by the null hypothesis $H_0$ and the alternative hypothesis $H_1$, respectively, we use the notation in \eqref{Ht} because the methodology is not limited to testing the traditional null and alternative hypothesis.

Based on the analysis of \cite{leifeld2012information}, a $p$-value was observed that fell above the significance threshold of .05 resulting in non-significant effect. A non-significant effect, however, does not imply there is evidence that the effect is zero because $p$-values can only be used for falsifying a null hypothesis. Furthermore, a non-significant finding may be a consequence of an underpowered study. Thus, based on the reported non-significant effect, it is unclear whether there is support for a zero effect of preference similarity or whether the data are inconclusive about which hypothesis is supported by the data.

A secondary question of interest was a specific ordering of effect sizes for different theoretical elements. \citet{leifeld2012information} were primarily interested in opportunity structures, such as shared committees, followed by influence attribution, followed by preference similarity. We will explore in Section~\ref{section_numerical} whether this theoretical importance ranking of effects translates into an order of effect sizes by testing this hypothesis against its equality-constrained counterpart, and against the complement:
\begin{eqnarray}
\nonumber H_1&:&\beta_{\text{committee}} > \beta_{\text{infl att}} > \beta_{\text{pref sim}}\\
\label{test2_AppA} H_2&:&\beta_{\text{committee}} = \beta_{\text{infl att}} = \beta_{\text{pref sim}}\\
\nonumber H_3&:& \text{``neither $H_1$, nor $H_2$.''}
\end{eqnarray}

\subsection*{Application B: Institutional and functional overlap in co-\-author\-ship networks}
The second motivating application is the nesting of co-authorship ties in institutional settings. \citet{leifeld2018polarization} used an ERGM to model the co-authorship networks among all political scientists in Germany and Switzerland during a five-year period. Of particular interest in the analysis was the question whether nomological and idiographic researchers, with their different publication norms, would find themselves separated into distinct communities in the co-authorship network. To answer this question, the study also took into account institutional incentives to collaborate: The more functional overlap existed between any two researchers, the greater the expected odds of collaboration between them, ranging from being in the same discipline and country, working at the same institution, being on the same team, all the way down to direct supervision relationships. But due to the lack of a nested hypothesis testing framework, the original study formulated the effect of increasing levels of institutional and functional overlap as three separate hypotheses, one for each level of functional overlap, without regard for the nested nature of the hypotheses. It concluded from increasingly positive coefficients that more functional overlap must lead to greater odds of collaboration. Here, we evaluate the nested nature of expected collaboration vis-a-vis alternative hypotheses stating that all levels of functional overlap have identical positive effects on co-authorship tie formation and that none of the different functional overlaps exerts any effect on tie formation.

\paragraph{$\textbf{H}_1$: Increasingly positive effect of functional overlap.}
The first hypothesis posits that researchers with increasing functional overlap collaborate at higher rates: Researchers with affiliations at the same institution (e.\,g., university or research institute) are hypothesized to have greater odds of forming collaboration ties than researchers without a shared affiliation. Researchers were expected to form ties at even higher rates within teams, such as chair groups or junior research groups, than within institutions, compared to pairs of researchers without a shared affiliation. Within teams, researcher pairs where one of the researchers was a professor (usually assistant, associate, or full professor) and the other one was not a professor (usually postdoctoral researcher or PhD student) were expected to have greater odds of co-authorship than the wider team they were embedded in, relative to the reference group of not sharing an affiliation. Mathematically, the hypothesis can be formulated as
\[
H_1:\beta_{\text{supervision}}>\beta_{\text{same team}}>\beta_{\text{same affiliation}}>0
\]
Like in the original study, we control for the number of edges in the network, having shared partners (GWESP with a decay of 0.3), degree distribution (GWDegree with a decay of 0.4), seniority, gender, gender homophily, geographic distance, topic similarity, the share of English articles among an author's publications, the number of publications, and the similarity between two researchers in their share of English article. Given the notation of a constrained hypothesis in \eqref{Ht}, the vector of coefficients is given by
\[
\begin{aligned}
\bm\beta  = (& \beta_{\text{edges}},\beta_{\text{supervision}},\beta_{\text{same team}},\beta_{\text{same affiliation}}, \beta_{\text{GWESP}},\beta_{\text{GWDegree}},\beta_{\text{seniority}},\beta_{\text{gender}}, \\
             & \beta_{\text{gender homophily}},\beta_{\text{geographic distance}},\beta_{\text{topic similarity}},
             \beta_{\text{share Eng. art.}},\beta_{\text{number publications}},\\
&\beta_{\text{Eng. art. similarity}})^{\top}.
\end{aligned}
\]
Thus, hypothesis $H_1:\beta_{\text{supervision}}>\beta_{\text{same team}}>\beta_{\text{same affiliation}}>0$ in Application B can be written using the matrix notation in \eqref{Ht} as
\[
\textbf{R}_1^O = \left[
\begin{array}{ccccccccccccc}
0 & 1 & -1 & 0 & 0 & \cdots & 0\\
0 & 0 & 1 & -1 & 0 & \cdots & 0\\
0 & 0 & 0 & 1 & 0 & \cdots & 0
\end{array}
\right]
\]

\paragraph{$\textbf{H}_2$: Positive and constant effect of functional overlap.}
The second hypothesis posits that having the same affiliation is the only (positive) factor that plays a role in the tendency to publish together. This hypothesis suggests that affiliations serve as large collaboration communities where there is no additional collaboration effect for researchers who form a supervisor--supervisee pair or for researchers who work on the same team (and who are not in a supervisor--supervisee pair). Mathematically, the hypothesis can be formulated as
\[
H_2:\beta_{\text{supervision}}=\beta_{\text{same team}}=\beta_{\text{same affiliation}}>0
\]
We control for the same covariates as under $H_1$. The matrix of equality constraints under this hypothesis is then given by
\[
\textbf{R}_2^E = \left[
\begin{array}{ccccccccccccc}
0 & 1 & -1 & 0 & 0 & \cdots & 0\\
0 & 0 & 1 & -1 & 0 & \cdots & 0
\end{array}
\right],
\]
and the matrix for the one-sided constraint under this hypothesis is equal to
\[
\textbf{R}_2^O = \left[
\begin{array}{ccccccccccccc}
0 & 0 & 0 & 1 & 0 & \cdots & 0
\end{array}
\right].
\]
Note that the same hypothesis would be formulated if $\textbf{R}_2^O$ had a one at the second or third column, and zero elsewhere, because the second, third, and fourth effect are restricted to be equal in $\textbf{R}_2^E$.

\paragraph{$\textbf{H}_3$: No Functional overlap effect.}
The third hypothesis assumes there is no effect of any level of functional overlap, which implies that researchers who form a supervisor--supervisee pair, researchers on the same team (and who do not form a supervisor--supervisee pair), researchers having the same affiliation (and who are not members of the same team), and researchers who do not have the same affiliation all have the same average tendency to publish together. Mathematically, the hypothesis can be formulated as
\[
H_3:\beta_{\text{supervision}}=\beta_{\text{same team}}=\beta_{\text{same affiliation}}=0
\]
We control for the same covariates as under $H_1$. The matrix of equality constraints under this hypothesis is then given by
\[
\textbf{R}_3^E = \left[
\begin{array}{ccccccccccccc}
0 & 1 & -1 & 0 & 0 & \cdots & 0\\
0 & 0 & 1 & -1 & 0 & \cdots & 0\\
0 & 0 & 0 & 1 & 0 & \cdots & 0
\end{array}
\right].
\]

\paragraph{$\textbf{H}_4$: Complement.}
The fourth, complement hypothesis that none of the assumptions under $H_1$, $H_2$, or $H_3$ is true. We write this hypothesis as
\[
H_4:\text{neither $H_1$, $H_2$, nor $H_3$.}
\]
We control for the same covariates as under $H_1$. Note that hypothesis $H_4$ covers the parameter space of the coefficients $\bm\beta$ that does not satisfy any of the constraints under $H_1$, $H_2$, and $H_3$.


\section{Bayesian hypothesis testing}
\subsection{Background information}
\begin{figure}[t]
\begin{center}
\includegraphics[height=3cm]{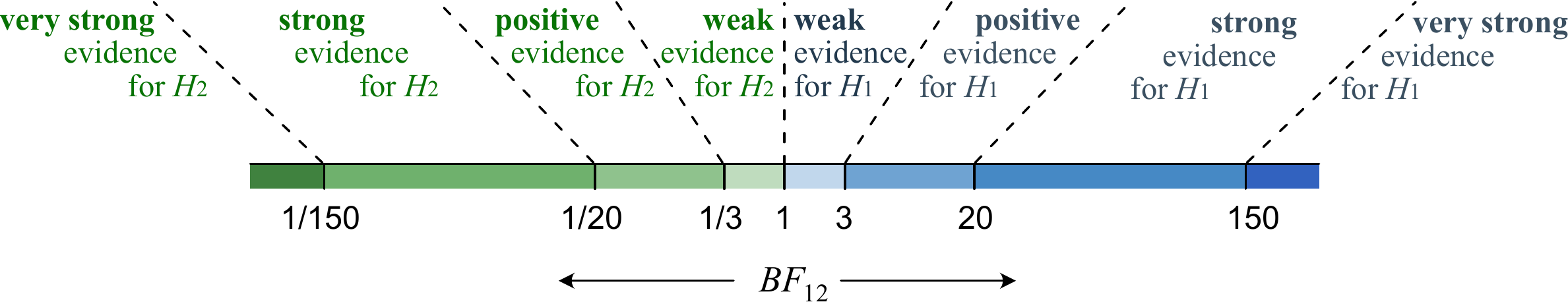}
\caption{Guidelines for interpreting the Bayes factor $BF_{12}$ of hypothesis $H_1$ against $H_2$ based on \cite{Raftery:1995}.}
\label{BF_line}
\end{center}
\end{figure}

The proposed Bayesian hypothesis test is based on the marginal (or integrated) likelihood of the observed network under hypothesis $H_t$, which is defined by
\begin{equation}
p(\textbf{Y}|\textbf{X},H_t) = \int p_t(\textbf{Y}|\bm\beta_t,\textbf{X}) p_t(\bm\beta_t) d \bm\beta_t,
\end{equation}
where $\bm\beta_t$ denotes the vector of the free parameters under $H_t$, and $p_t(\bm\beta_t)$ denotes the prior probability distribution of the free parameters under $H_t$, which reflects the prior expectation and uncertainty about the free parameters before observing the data. Under an ERGM, the marginal likelihood can be computed using the algorithm of \cite{bouranis2018bayesian}. If we consider two competing hypotheses, the ratio of the respective marginal likelihoods is also known as the Bayes factor. For example, the Bayes factor between hypothesis $H_1$ and $H_2$ is defined by
\begin{equation}
BF_{12} = \frac{p(\textbf{Y}|\textbf{X},H_1)}
{p(\textbf{Y}|\textbf{X},H_2)},
\end{equation}
which quantifies how much more plausible the observed network is under $H_1$ relative to $H_2$. From this definition, it automatically follows that the data were more likely to be observed under $H_1$ than $H_2$ if $BF_{12}>1$ and the data were more likely to be observed under $H_2$ if $BF_{12}<1$. Because of this interpretation, the Bayes factor can be used as a relative measure of evidence in the data in favor of one hypothesis against another hypothesis. For example, if $BF_{12}=10$ for the hypotheses in Application A, this implies that there is 10 times more evidence in the data for hypothesis $H_1$ than for the alternative $H_2$. Furthermore, it can be seen that Bayes factors satisfy a symmetry property, i.\,e., $BF_{21}=1/BF_{12}$, implying that if there is 10 times more evidence for $H_1$ relative to $H_2$, then there is 10 times less evidence in the data for $H_2$ relative to $H_1$. Thus, none of the hypotheses has a special role (unlike the null hypothesis in significance testing). Moreover, the Bayes factor follows a transitivity property, e.\,g., $BF_{31}=BF_{32}\times BF_{21}$, which implies that if $H_3$ receives 5 times more evidence than $H_2$, and $H_2$ receives 10 times more evidence than $H_1$, then $H_3$ receives 50 times more evidence than $H_1$. To facilitate the interpretation, Figure \ref{BF_line} shows the continuous line of the evidence between two hypotheses as quantified by the Bayes factor, together with a discretized categorization \citep{Raftery:1995}. Note that this categorization does not have to be applied in a strict sense (as the evidence lies on a continuous scale) but it can be useful for researchers who are unfamiliar with Bayes factors. Finally, we note here that Bayes factors are consistent under mild conditions \citep[e.\,g.][]{OHagan:1995}, which implies that the evidence for the true hypothesis goes to infinity as the sample size grows. Thus, $BF_{12}\rightarrow \infty$ as $N\rightarrow\infty$ if $H_1$ is true, and
$BF_{12}\rightarrow 0$ as $N\rightarrow\infty$ if $H_2$ is true.


Furthermore, in a Bayesian framework, one can specify prior probabilities for the hypotheses, denoted by $P(H_t)$ for $t=1,\ldots,T$, which reflect the plausibility of the hypotheses (such as scientific expectations or substantive theories) before observing the data. A common default choice is the use of equal prior probabilities, i.\,e., $P(H_t)=T^{-1}$ for $t=1,\ldots,T$. Subsequently, when data are observed, the evidence between the hypotheses in the data is quantified in the Bayes factor, which can be used to update the prior odds of the hypotheses to obtain the posterior odds via
\begin{equation}
\frac{P(H_1 | \textbf{Y},\textbf{X})}
{P(H_2 | \textbf{Y},\textbf{X})} = BF_{12} \times \frac{P(H_1)}
{P(H_2 )}
\end{equation}
The posterior probabilities of the hypotheses can then be computed via
\begin{equation}
P(H_t | \textbf{Y},\textbf{X}) = \frac{P(H_t)BF_{t1}}{\sum_{t'=1}^TP(H_{t'})BF_{t'1}}.
\end{equation}

\subsection{Prior specification and Bayes factor computation}
In order to compute the marginal likelihoods of the hypotheses and, subsequently, the Bayes factors between the hypotheses, one only needs to formulate the prior distributions for the free parameters under the hypotheses. To simplify this endeavor, we use an encompassing prior approach \citep{Klugkist:2007} where an unconstrained (or encompassing) prior, denoted by $p_u(\bm\beta)$, is formulated under the full unconstrained ERGM, denoted by $H_u$. Truncations are used under the constrained hypotheses according to
\[
p_t(\bm\beta) \propto p_u(\bm\beta) \times I(\textbf{R}_t^E\bm\beta=\textbf{0} ~ \cap ~ \textbf{R}_t^O\bm\beta>\textbf{0}),
\]
where $I(\cdot)$ denotes the indicator function. 
Thus, instead of needing to formulate prior distributions under all $T$ hypotheses, we only need to formulate one unconstrained prior. Another advantage of this choice of the prior is that it substantially simplifies the computation of the Bayes factors as the marginal likelihoods do not have to be computed\footnote{Marginal likelihoods can be done using the \texttt{evicencePP} or \texttt{evidenceCJ} functions in the \texttt{Bergm} package \citep{bouranis2018bayesian,caimo2013bayesian}. Priors need to be carefully specified.} but instead only the prior and posterior distribution under the full unconstrained ERGM are needed. In fact, the Bayes factor between a constrained hypothesis $H_t$ and the unconstrained model can be written as an extended Savage-Dickey density ratio \cite[e.\,g.][]{Dickey:1971,Wetzels:2010,mulder2021bayes}:
\begin{equation}
BF_{tu} = \frac{p_u(\bm\beta^E_t=\textbf{0}|\textbf{Y},\textbf{X})}{p_u(\bm\beta^E_t=\textbf{0})}\times
\frac{\text{P}_u(\bm\beta^O_t>\textbf{0}|\bm\beta^E_t=\textbf{0},\textbf{Y},\textbf{X})}
{\text{P}_u(\bm\beta^O_t>\textbf{0}|\bm\beta^E_t=\textbf{0})},
\label{Btu}
\end{equation}
where $\bm\beta^E_t = \textbf{R}^E_t\bm\beta$ and $\bm\beta^O_t = \textbf{R}^O_t\bm\beta$. Thus, to compute the first term on the right hand side, we only need to compute the posterior and prior density of the transformed parameter $\bm\beta^E_t$ at the equality constrained null value. For the second term, we need to compute the posterior and prior probability that the order constraints hold under the condition that the equality constraints hold under the unconstrained model. Note that the posterior density at the null value and the (conditional) posterior probability that order constraints hold serve as measures of relative fit of the constrained hypothesis (relative to the unconstrained model). The prior density at the null value and the prior probability that the order constraints hold serve as measures of relative complexity of the constrained hypotheses (relative to the unconstrained model). This explicitly shows how the Bayes factor functions as an Occam's razor by balancing between fit and complexity when quantifying the relative evidence in the data \citep[e.g.,][]{hoijtink2019tutorial}.


An unconstrained multivariate normal prior distribution is specified for the ERGM coefficients, which is inspired by the $g$-prior approach for linear and logistic regression models \citep{Zellner:1986,Liang:2008,hanson2014informative}. The proposed prior can be used in an automatic fashion without requiring manual specification of the prior hyperparameters using external knowledge or simulation techniques. First, we set a noninformative flat prior for the edges parameter as is common for the intercept in the $g$-prior. This choice ensures that the number of ties in the network is not shrunk towards any (subjectively specified) prior mean. A consequence is that the edges parameter cannot be tested because improper flat (or arbitrarily vague) priors cannot be used for hypothesis testing using Bayes factors \citep{OHagan:1995,Berger:1996}. This is acceptable as the edges parameter typically serves as a nuisance parameter in an ERGM. We can implement it as a normal prior for the edges coefficient using a mean of 0 and a very large variance (e.g., 1e4). Furthermore, the prior means of the other coefficients are also set to zero. This has two important motivations. First, a prior mean of 0 implies that, a priori, values close to zero are more likely than values far away from zero which implies that smaller effects are on average more likely than (very) large effects \citep{Jeffreys}. Second, as the prior is symmetrical around 0 negative values are equally likely as positive values, which is an objective prior choice for one-sided testing \citep{Mulder:2010}. Furthermore, following the $g$-prior approach, the prior covariance matrix of the ERGM coefficients is based on the implied covariance structure of the predictor variables that is rescaled to one dyadic observation, i.\,e., $D(\bm\Delta^{\top}\bm\Delta)^{-1}$, where $D$ is the total number of possible dyads in the network and $\bm\Delta$ is the design matrix of change statistics (excluding the ones for the edges covariate) under the full ERGM, which is constructed by stacking the row vectors of change statistics $\bm\delta_{(ij)}(\textbf{Y},\textbf{X})^{\top}$ in Equation \eqref{changestats} over all dyads $(i,j)$. This results in a weakly informative prior that is dominated by the information in the observed network.
Hence, the unconstrained prior can be formulated as
\begin{equation}
p_u(\bm\beta|\textbf{X}) = N\left(\textbf{0},\text{diag}\left(1\text{e}4,D(\bm\Delta^{\top}\bm\Delta)^{-1}\right)\right).
\label{normalprior}
\end{equation}

Given the multivariate normal prior for the ERGM parameters, a sample from the unconstrained posterior can obtained in a straightforward manner using the \texttt{Bergm} package \cite{caimo2014bergm}. Following large sample theory, a posterior can often be well approximated using a Gaussian approximation \citep[][]{gelman2013bayesian,yin2022highly}. Relatedly, note that $p$-values under ERGMs can also be computed using a Wald-type test by relying on Gaussian approximations. We approximate the unconstrained posterior of the ERGM parameters with a multivariate Gaussian distribution, i.e.,
\begin{equation}
p_u(\bm\beta|\textbf{X},\textbf{Y}) \approx N(\bar{\bm\beta},\bar{\bm\Sigma}),
\label{normalpost}
\end{equation}
where $\bar{\bm\beta}$ and $\bar{\bm\Sigma}$ denote the approximated posterior mean and the approximated posterior variance, respectively. Using this approximation, the posterior density and posterior probability in \eqref{Btu} is straightforward, and Bayes factors between the hypotheses of interest can easily be obtained.


\subsection{Software implementation in \texttt{BFpack}}
The proposed methodology for testing hypotheses using the Bayes factor is implemented in the \texttt{R} package \texttt{BFpack} \citep[][version 1.2.3 or later]{mulder2021bfpack}\footnote{Version 1.2.3 of \texttt{BFpack} can be installed from CRAN by running: \texttt{install.packages("BFpack")}.}. This package contains a variety of Bayes factor testing procedures for equality/order-constrained hypotheses for different types of models, including (but not limited to) (multivariate) multiple regression, (multivariate) analysis of (co)variance, generalized linear models, correlational analysis. The main function of the package is \texttt{BF()}. It only requires a fitted model in the form of a \texttt{R} object. Depending on the class of the object, Bayes factors and posterior probabilities are computed under the framework of the fitted model. As part of the current paper, the package is extended to also support objects of class \texttt{ergm} or \texttt{bergm}. Finally, hypotheses with equality and order constraints on key ERGM coefficients can be formulated using a character string specified in the \texttt{hypothesis} argument.

To compute the Bayes factors and posterior probabilities among a set of competing hypotheses under an ERGM, a fitted ERGM object of class \texttt{ergm} \citep[from the \texttt{ergm} package;][]{hunter2008ergm,krivitsky2023ergm} or a fitted Bayesian ERGM of class \texttt{bergm}\footnote{\texttt{BF} can only run on a \texttt{bergm} object when the network data object is also loaded in the environment because a \texttt{bergm} object does not contain the data object, which is required for the analysis.} \citep[from the \texttt{Bergm} package;][]{caimo2014bergm} needs to be plugged in the \texttt{BF} function. The same analysis is performed for an object of class \texttt{ergm} or \texttt{bergm} if the same ERGM is specified using the `formula' argument in these respective functions. The function then extracts the necessary output to compute the Bayes factors and posterior probabilities of interest. Note that when a \texttt{bergm} object is plugged in the function (regardless of the prior that was used to obtain this object), a new Bayesian ERGM is fitted using the proposed unit-information prior from which the Bayes factors and posterior probabilities are computed. To get accurate estimates of the Bayes factors and posterior probabilities, it is recommend to collect a sufficient number of posterior draws using the \texttt{bergm} function. Throughout this paper we set \texttt{main.iter = 100000} \cite[instead of the default setting of 1000 draws in the \texttt{bergm} function,][]{caimo2014bergm}, which can also be included as an argument in the \texttt{BF} function which is then transferred as argument for the \texttt{bergm} function.

In addition to the confirmatory hypothesis test of multiple constrained hypotheses of the form \eqref{Ht}, \texttt{BFpack} also computes the posterior probabilities of a zero effect, a negative effect, or a positive effect by default, i.e.,
\[
H_1:\beta_k = 0 \text{ vs }H_2:\beta_k < 0 \text{ vs }H_3:\beta_k > 0,
\]
for all coefficients $\beta_1,\ldots,\beta_K$ using equal prior probabilities. The output of these tests can be used for testing the direction of the individual effects on the formation of ties in the network. In Section 5, the software was used to test the empirically informed hypotheses from Section 2. 

\section{Numerical behavior} \label{section_numerical}
Numerical simulations were carried out to study the behavior of the proposed Bayesian test. First, a simulation was executed based on the \texttt{florentine} data \citep{Breiger1986} to illustrate a key difference between the behavior of the proposed Bayesian test and the classical $p$-value as a function of the network size. Also, a comparison is made with the BIC and the AIC. Second, simulations were executed to investigate the selection behavior in settings similar to the policy network model of \cite{leifeld2012information} with a simplified model specification using predictor variables: edges, preference similarity, influence attribution, shared committees, reciprocity, and geometrically weighted edgewise shared partner. The goals of this second set of simulations were (i) to explore the behavior when testing a null hypothesis and when testing an ordered hypothesis as a function of different true population values and (ii) to illustrate the consistent behavior of the test where the posterior probability of the true hypothesis gradually goes to 1 as the network size grows.

\subsection{Illustrating the behavior in comparison to classical $p$-values}
We first illustrate that a Bayes factor behaves differently in comparison to a $p$-value as a function of the sample (or network) size. The difference in behavior is a consequence of the fact that the Bayes factor is statistically consistent if the simpler (null) model is true while the $p$-value is not. As the sample (or network) size grows to infinity, the evidence for a true null model as quantified by the Bayes factor grows to infinity (and correspondingly, the posterior probability of the true null goes to 1). A significance test based the $p$-value on the other hand is statistically inconsistent. There always is a probability (typically .05) to incorrectly reject the null, even when the network size grows to infinity. To illustrate the different behavior, we consider a relatively small network that results in a 2-sided $p$-value of .05 for an effect of a predictor variable. Next we consider larger networks in such a way that the $p$-value for this effect remains constant at .05. This is achieved by letting the true parameter value gradually decrease to zero, and thus the null is true in the limit. Consequently, the evidence for the null using the Bayes factor then gradually increases, eventually towards infinity. In the literature, this conflicting behavior is also referred to as the Jeffreys-Lindley paradox \citep{Jeffreys,Lindley:1957}.

We consider the following simple ERGM based on the Florentine data \citep{Breiger1986}: 
\begin{verbatim}
flomarriage ~ edges + kstar(2) + absdiff("wealth")
\end{verbatim}
and the hypothesis test of whether the effect of the absolute difference in wealth, \texttt{absdiff("wealth")}, is zero or not, i.e.,
\begin{eqnarray*}
H_1&:&\beta_{\text{absdiff wealth}}=0\\
H_2&:&\beta_{\text{absdiff wealth}}\not=0.
\end{eqnarray*}
Simulated networks are considered consisting of $n = 7, 10, 15, 20, ..., 50,$ and $55$ actors. The parameter values of the \texttt{edges} and \texttt{kstar(2)} statistic were set equal to the MLEs of the empirical florentine data (although the exact choices did not qualitatively affect the results), and the effect of \texttt{absdiff("wealth")} was set such that the $p$-values of the generated data was close to .05. The covariate values of the actors were sampled from the empirical \texttt{wealth} values with replacement. Of the generated networks, only the ones that resulted in a two-sided $p$-value in the interval (0.045,0.055) were stored, resulting in 300 networks for each network size $n$.

\begin{figure}[t]
\begin{center}
\includegraphics[height=10cm]{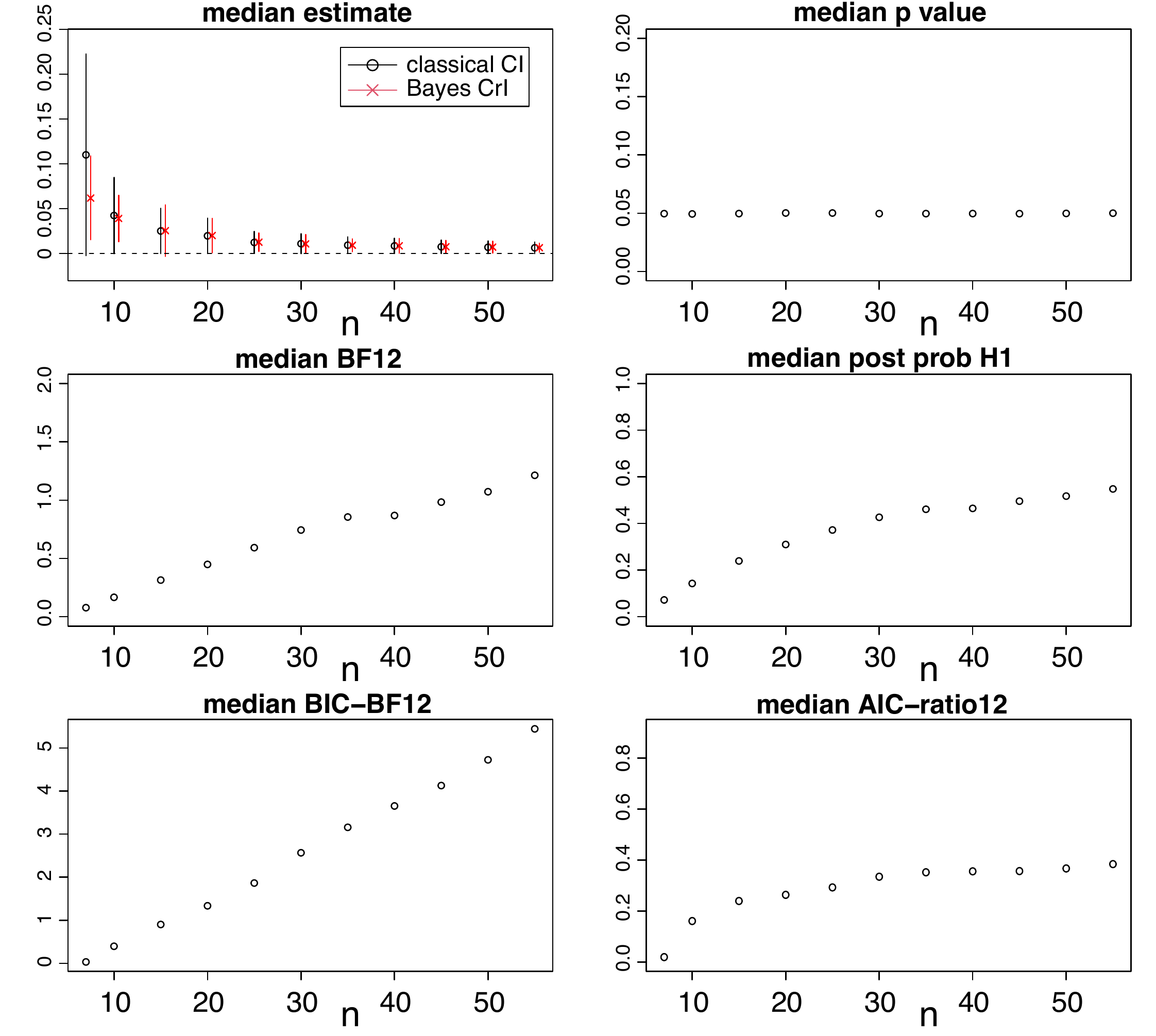}
\caption{The median of the estimated effect MLE and posterior mean with confidence (CI) and credible (CrI) bounds, respectively, the median of the $p$-value, the median of the Bayes factor of $H_1:\beta_{\text{absdiff("wealth")}}=0$ against $H_2:\beta_{\text{absdiff("wealth")}}\not=0$, the median of the posterior probability of $H_1$, the median of the evidence based on the BIC, and the median of the evidence based on the AIC based on 300 generated data sets as a function of the network size $n=7,~10,~15,~20,\ldots,$ and $55$.}
\label{new_simulation}
\end{center}
\end{figure}

Figure \ref{new_simulation} shows the median MLE (black circle) with with confidence bounds and the Bayesian posterior modes (red crosses) with credibility interval bounds (upper left panel), the median of the $p$-values (upper right panel), which is .05 by construction, the median of the Bayes factors of the simple hypothesis $H_1$ against the alternative $H_2$ (middle left panel), the median of the posterior probability of $H_1$ (assuming equal prior probabilities for $H_1$ and $H_2$) (middle right panel), the median of the approximated Bayes factor based on the BIC (lower left panel), and the median of the evidence ratio of $H_1$ against $H_2$ based on the AIC\footnote{The evidence based on the BIC and AIC were computed according to $\exp(-(BIC(H_1)-BIC(H_2))/2)$ and $\exp(-(AIC(H_1)-AIC(H_2))/2)$, respectively.} (lower right panel), all as a function of the network size. As the effect gradually decreases towards zero, we see that the evidence for the simpler hypothesis, which assumes that the effect is zero (middle left panel), gradually increases. Eventually it results in evidence in favor of the simpler hypothesis $H_1$ (around $n=50$) while the $p$-value remains constant at .05 (by construction). The posterior probability of $H_1$ shows similar behavior as the Bayes factor as it behaves as a normalized version of the evidence (middle right panel). The behavior of the Bayes factor is a consequence of its consistency, which implies that the evidence goes to infinity towards the true hypothesis as the network size grows. The $p$-value, on the other hand, is inconsistent (by definition) because it is uniformly distributed under the null regardless of the sample size. For this reason, the $p$-value also cannot be used as a quantification of the evidence in favor of the null \citep[see also][]{Rouder:2009}. The Bayes factor does not have this property as all hypotheses are compared on equal footing. If the evidence clearly points towards $H_1$ (or $H_2$), this implies that $H_1$ (or $H_2$) was most likely to have generated the data, and if the Bayes factor does not clearly point towards any of the hypotheses (i.e., the Bayes factor is close to 1), then there is absence of evidence and more data are required to obtain clearer evidence towards the hypothesis that is most likely true. This interpretation holds regardless of the network size.

As a comparison, we also provide the evidence ratios based on the BIC (Figure \ref{new_simulation}, lower left panel) and the AIC (Figure \ref{new_simulation}, lower right panel). Both show increasing trends as a function of the network size, similar to the proposed Bayes factor. However the evidence for the null based on the BIC increases considerably faster where an evidence ratio of approximately 1 is already obtained around $n=15$. Thus the BIC is considerably more conservative towards the null than the proposed Bayes factor. Note that the BIC is also sometimes regarded as being too conservative towards simpler models \citep[e.g.][]{weakliem1999critique}. To understand the difference in behavior, note that the proposed Bayes factor uses a unit information prior that is centered at 0 while the BIC behaves as a marginal likelihood using a unit information prior that is centered at the MLE \citep{kass1995reference,Raftery:1995}. Another difference is that the BIC is based on a Laplace approximation of the likelihood while the proposed Bayes factor is based on a Laplace approximation of a (kernel of the) posterior, which combines information in the unit information prior and the likelihood. We refer the interested reader to \cite{Raftery:1995}, \cite{weakliem1999critique}, and \cite{raftery1999bayes} for relevant discussions about the BIC approximation for hypothesis testing and model selection. Finally, we note that the evidence based on the AIC increases with a slower rate than the proposed Bayes factor, making the AIC more liberal towards larger models. This is related to the property of the evidence ratio based on the AIC which converges to a constant and not infinity \citep{OHagan:1995}, which implies inconsistent behavior in the long run as the sample size grows.

\subsection{Illustrating the behavior using simulated policy networks}
In this section, two hypothesis tests are discussed based on the empirical networks from Section~\ref{section_ergm}. First, a traditional null hypothesis test is considered of whether the coefficient of preference similarity equals zero or not, i.\,e.,
\[
\begin{array}{l}
H_1:\beta_{\text{pref sim}} = 0\\
H_2:\beta_{\text{pref sim}} \not= 0.
\end{array}
\]
Second, a multiple order-constrained hypothesis test is considered of a specific ordering of three coefficients versus an equality-constrained alternative versus the complement, i.\,e.,
\[
\begin{array}{l}
H_1:\beta_{\text{pref sim}} < \beta_{\text{infl attr}} < \beta_{\text{committee}}\\
H_2:\beta_{\text{pref sim}} = \beta_{\text{infl attr}} = \beta_{\text{committee}}\\
H_3: \text{neither $H_1$, nor $H_2$}.
\end{array}
\]

Networks were considered of 10 actors, 30 actors (as in the empirical data), and 90 actors. For the network of 30 actors, the population values were set equal to the MLEs of the fitted model to the empirical data. For the network of 10 actors, the population values were set equal to the MLEs based on a subset of the first 10 actors of the empirical data. For the network of 90 actors, the population values were equal to the MLEs based on the $90\times 90$ block diagonal adjacency matrix where the $30\times 30$ empirical adjacency matrix was placed 3 times on the block diagonal. This setting ensured that the mean degree of the simulated networks was comparable with the empirical network.

To investigate the behavior of the proposed test, only the values of the key parameters that were tested were changed. In the simulation for the first test, the value for $\beta_{\text{pref sim}}$ was varied on a grid from $-0.8$ to $0.8$. For the second test, network data were generated using values for $(\beta_{\text{pref sim}}, \beta_{\text{infl attr}}, \beta_{\text{committee}})$ equal to $(\beta,2\beta,3\beta)$, where $\beta$ was varied on a grid of $-0.4$ to 0.4. Thus, when $\beta$ is positive, the order-constrained hypothesis $H_1$ is true, when $\beta$ is zero the equality hypothesis $H_2$ is true, and when $\beta$ is negative, the complement hypothesis $H_3$ is true. For each grid value, 100 networks were generated, and Bayes factors were computed between the hypotheses and transformed to posterior probabilities. We checked whether the posteriors can be accurately approximated with Gaussian distributions. This was the case over the entire range of values for the tested parameters. The plots looked similar as in Figure \ref{fig_AppA_approx} and \ref{fig_AppBF_approx} (discussed later for the empirical applications), and therefore the similarly looking plots for the simulated data are omitted here.

\begin{figure}[t]
\begin{center}
\includegraphics[height=10cm, width=12cm]{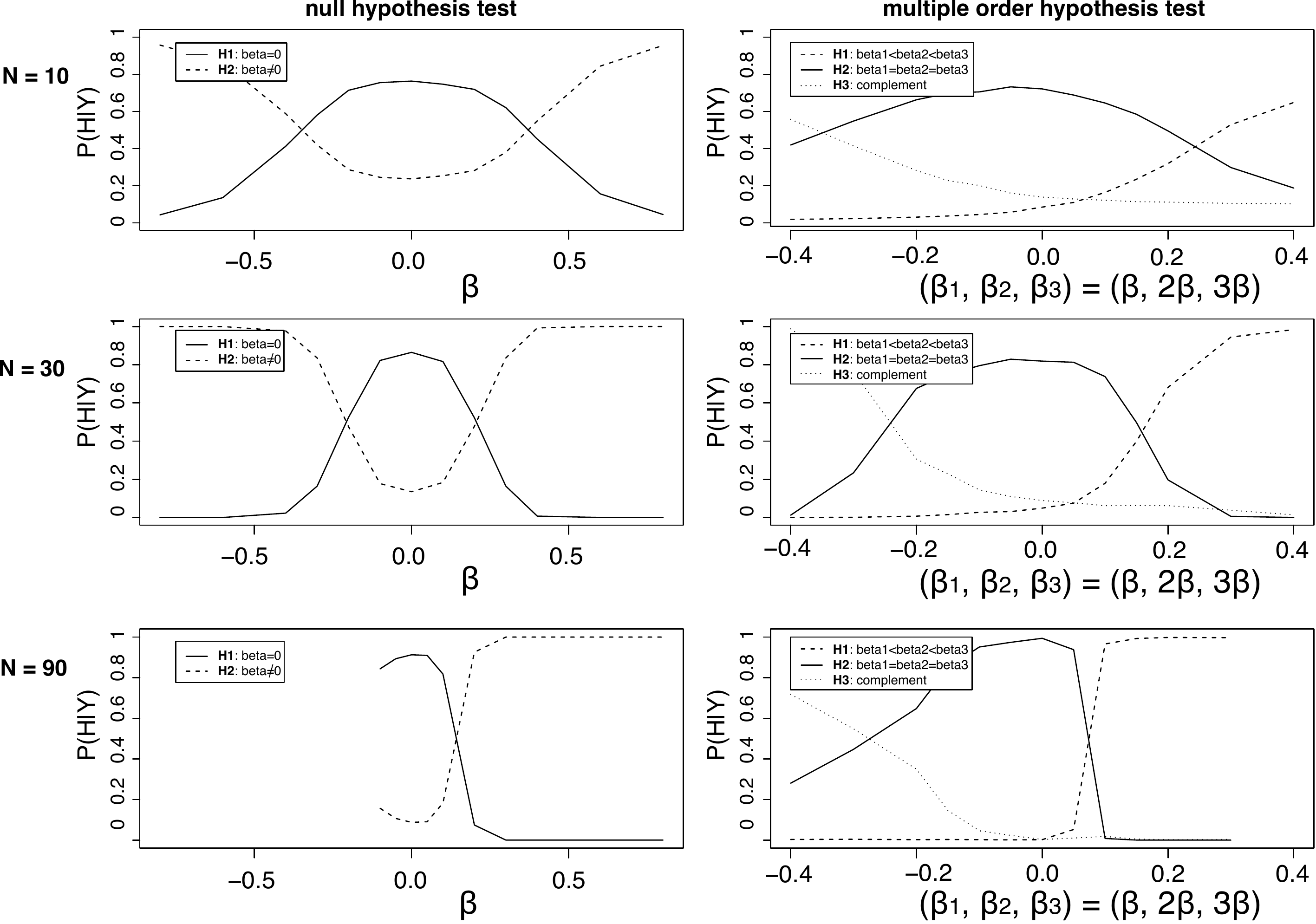}
\caption{Median posterior probability of the first precise test (left panels; $H_1:\beta=0$ vs $H_2:\beta\not=0$) and the multiple order test (right panels; $H_1:\beta_1<\beta_2<\beta_3$ vs $H_2:\beta_1=\beta_2=\beta_3$ vs $H_3:$ complement) for networks of 10 actors (upper panels), 30 actors (middle panels), and 90 actors (lower panels) based on 100 randomly generated networks for every grid value. For the generated networks with 90 actors, the ERGM could often not be fitted for negative grid values for both tests. For the first test the lines are therefore omitted.}
\label{fig_sim}
\end{center}
\end{figure}

Figure \ref{fig_sim} shows the median posterior probability of the first test (left panels) and the multiple order test (right panels) for networks of 10 actors (lower panels), 30 actors (middle panels), and 90 actors (lower panels). Before discussing the results, we note that for the larger networks with 90 actors, the MLE often did not exist in the lower range of the parameter space in both tests. This seemed to be caused by the generated networks being too sparse. For the first test, the results are therefore omitted for parameter values smaller than $-0.2$, and in the second test, only the generated networks for which the MLE existed were used, resulting in more moderate estimates (and, thus, more moderate outcomes of the evidence).

For both tests, the figure shows the anticipated behavior, where on average the true hypothesis receives the largest posterior probability and this increases for larger networks. Furthermore, we see that the posterior probability for the equality hypothesis ($H_1$ in the first test and $H_2$ in the second test) goes to 1 with a slower rate when it is true than the rate when the alternative hypotheses are true. This implies that more data are generally needed to get a large posterior probability of a true equality hypothesis than for unconstrained or order-constrained hypotheses. Intuitively, this seems reasonable because it is easier to find evidence that a parameter is not exactly 0 (or that several parameters are not exactly equal) than to find evidence that a parameter is \textit{exactly} 0 (or that several parameters are \textit{exactly} equal). This behavior is also often observed for Bayes factors for other testing problems.

For the order-constrained hypothesis test, it is interesting to see that smaller effects are needed in the direction of the order-constrained hypothesis to acquire the same posterior probabilities in comparison to the effect in the opposite direction for the complement hypothesis. For example, in the case of a network with $N=30$ actors (right middle panel of Figure \ref{fig_sim}), the posterior probability for the order-constrained hypothesis $H_1$ is largest when the true value is about 0.16 while the posterior probability for the complement hypothesis $H_3$ is largest when the effect $\beta$ is lower than $-0.24$. This can be explained by the fact that the order-constrained hypothesis covers a smaller parameter space than the complement hypothesis, and therefore the complement receives a larger penalty for model complexity. In fact, the complexity of order-constrained hypotheses is implemented via the prior probability that the constraints hold \citep[e.\,g., see][]{Mulder:2010}. This illustrates that the criterion also behaves as an Occam's razor when assessing order-constrained hypotheses.

\begin{figure}[t]
\begin{center}
\includegraphics[height=6cm]{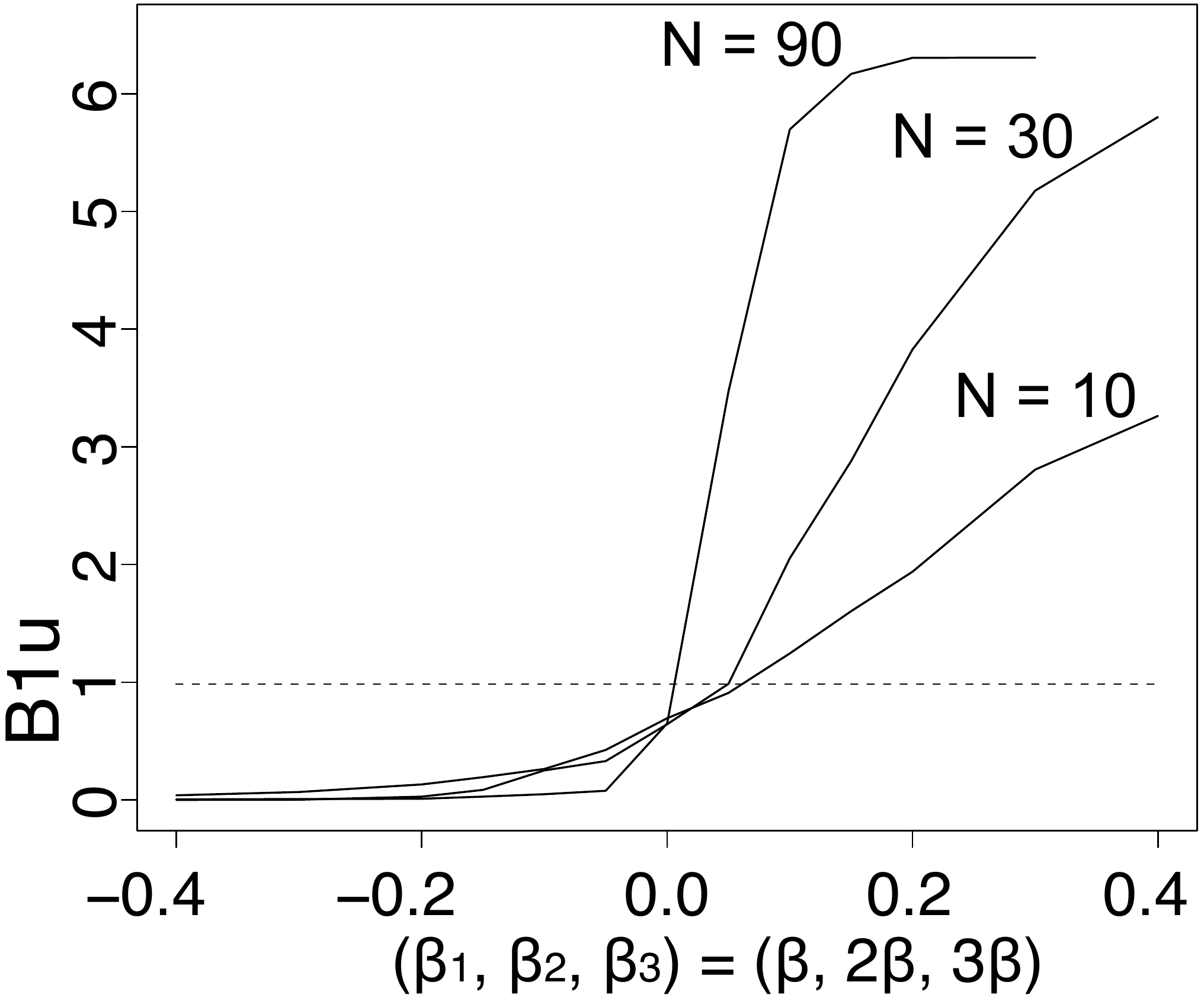}
\caption{The median Bayes factor of the order-constrained hypothesis $H_1:\beta_{\text{pref sim}} < \beta_{\text{infl att}} < \beta_{\text{committee}}$ against the unconstrained model, which is computed as the ratio of posterior and prior probabilities that the constraints hold, while varying the ordered effect size on the $x$ axis, for different network sizes of $N=10$, 30, and 90 actors.}
\label{fig2_sim}
\end{center}
\end{figure}

To see this more clearly, Figure \ref{fig2_sim} displays the Bayes factors of order-constrained hypothesis $H_1$ versus the unconstrained model, which is computed as the posterior probability that the order constraints hold divided by the prior probability that the order constraints hold under the unconstrained model; see Equation \eqref{Btu}. Thus, in the case of overwhelming evidence for the order constraints, the posterior probability that the order constraints hold is approximately 1, and the Bayes factor of the order-constrained hypothesis against the unconstrained model is approximately equal to the reciprocal of the prior probability that the constraints hold. Note that the prior probability depends on the prior covariance structure, which depends on the design matrix via \eqref{normalprior}. For example, if the covariates are independent, the prior probability that the order constraints $\beta_{\text{pref sim}} < \beta_{\text{infl attr}} < \beta_{\text{committee}}$ hold under the unconstrained model is equal to 1/6, which can be explained by the fact that there are 6 possible orderings of 3 parameters. In this case, the Bayes factor of the order-constrained hypothesis against the unconstrained model grows to 6 if the ordered effect increases. This is also what can be observed in Figure \ref{fig2_sim} for a slightly different prior covariance matrix \citep[based on the data from][]{leifeld2012information}.

\section{Empirical applications (revisited)} \label{section_app}
\subsection*{Application A: Information exchange in policy networks}
\begin{figure}[t]
\begin{center}
\includegraphics[width=8cm]{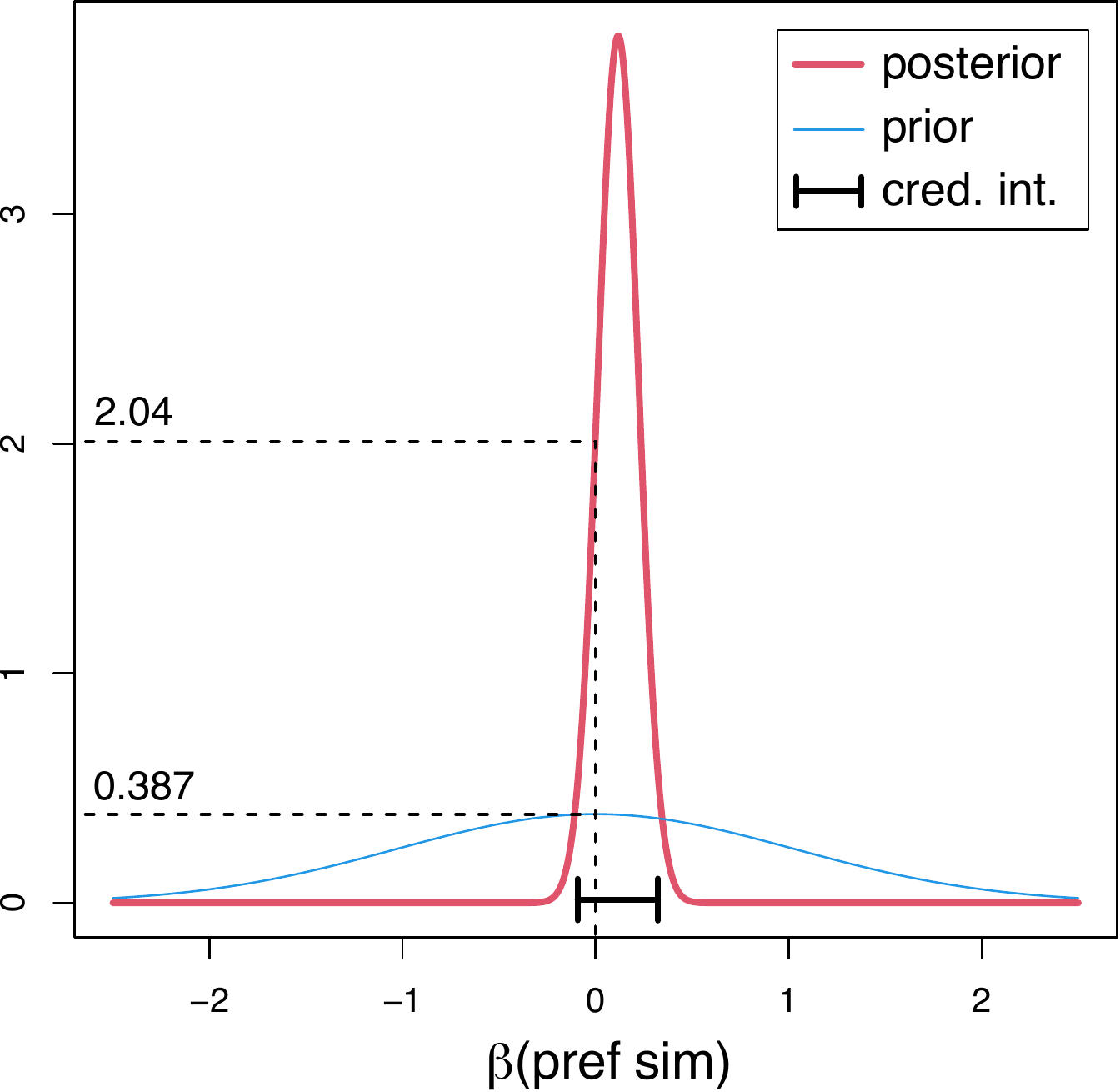}
\caption{Approximated posterior (thick red line) and prior (thin blue line) for the effect of preference similarity based on the policy network. The posterior and prior density at the null value are equal to 2.04 and 0.387, respectively, resulting in a Bayes factor of $\frac{2.04}{0.387}\approx 5.2$ in favor of a zero effect. The thick black line denotes the 95\% Bayesian credibility interval.}
\label{fig.post.prior}
\end{center}
\end{figure}

The main interest was whether preference similarity plays a role in information sharing in policy networks or whether the effect is absorbed by institutional, relational, and social opportunity structures, i.e.,
\[
\begin{array}{l}
H_1:\beta_{\text{pref sim}} = 0\\
H_2:\beta_{\text{pref sim}} \not= 0.
\end{array}
\]
These hypotheses are tested under Model 2 of \cite{leifeld2012information}\footnote{The \texttt{R} code for analyses can be found in Appendix \ref{AppA}.}. The Bayes factor between $H_1$ and $H_2$ was computed using Equation \eqref{Btu} (only using the first part on the right hand side) where the posterior and prior density at 0 are approximately equal to 2.04 and 0.387, respectively, resulting in $BF_{12} = \frac{2.04}{0.387} = 5.2$ (see Figure \ref{fig.post.prior}), which implies that the network was 5.20 times more likely to be generated under a model that assumes that preference similarity equals zero than by a model that assumes that preference similarly can attain any real value. This suggests positive evidence in favor of no effect of preference similarity on tie formation. Note that the computation of the evidence as the posterior and prior ratio at the null value shows an intuitive correspondence with Bayesian estimation: When the posterior density at the null value is high (low), the null value is likely (unlikely) according to the posterior and the evidence in favor of the null hypothesis value is also high (low) \citep[see also][]{Wagenmakers:2010}.

When assuming equal prior probabilities for the two hypotheses, i.e., $P(H_1)=P(H_2)=.5$, the posterior probabilities are then given by $P(H_1|\textbf{Y})=.839$ and $P(H_2|\textbf{Y})=.161$. Thus after observing the data, there is a posterior probability of approximately 84\% for the model that assumes that preference similarity plays no role in tie formation in this policy network. Note that no formal decision needs to be drawn based on these posterior probabilities. But if one would conclude that $H_1$ is the true hypothesis, there would be a conditional error probability of about 16\% of drawing the wrong conclusion given the observed network conditional on the observed network.

Further note again that the two-sided $p$-value of $.136$ in comparison to the traditional threshold value of .05 would imply that there was not enough evidence to reject the null. Thus, based on the $p$-value, we cannot tell whether there is evidence of no effect of preference similarity (because $p$-values cannot be used as a measure of support for a zero effect) or absence of evidence (i.e., the information in the data is inconclusive). The Bayesian test, on the other hand, removes this uncertainty as its outcome suggests that there is some positive evidence that the effect is zero, and thus, that the non-significant result is not a consequence of an underpowered study. The weight of evidence between $H_1$ and $H_2$ was also computed using the BIC and the AIC, which resulted in 13.63 (i.e., positive evidence for $H_1$) and 1.26 (i.e., practically equal evidence between the hypotheses with a very slight preference towards $H_1$), respectively. This result is in agreement with the simulation results presented in Section 4.1, which also showed that the BIC is generally more conservative towards the simpler model and the AIC is more liberal towards the larger alternative.

Besides testing the effect of preference similarity, we also discuss provide the results for the other coefficients. Table \ref{taBF_AppA} presents the MLEs, the standard errors, the $p$-values, the Bayesian posterior means, the posterior standard deviations (both based on the proposed unit information prior), and the posterior probabilities when testing whether a coefficient is zero, negative, or positive (assuming equal prior probabilities of $\frac{1}{3}$). The latter posterior probabilities are standard output of \texttt{BFpack}. Note that the Bayesian testing results are not provided for the (nuisance) edges coefficient because an improper flat prior is used for this parameter. When looking at the $p$-values and the posterior probabilities of the null hypotheses of the other coefficients in Table \ref{taBF_AppA_test2}, we see that generally both statistics point in the same direction. However the interpretation and theoretical properties of these methods (e.g., see the simulations in Section 4.1) are fundamentally different. This needs to be taken into account when interpreting these results. Furthermore, we can see that most (but not all) posterior means are closer to zero than the MLEs. This could be attributed to some prior shrinkage towards the prior mean of 0.

\begin{table}[t]
\begin{center}
\caption{Classical results (MLE, standard error, and $p$-value), Bayesian estimates (using the normal unit information prio), and posterior probabilities of a null, a negative, and a positive effect (assuming equal prior probabilities) for all ERGM coefficients of the policy network (Application A).}
\hspace*{-2.2cm}
{\small 
\begin{tabular}{lccccccccccc}
  \hline
&  \multicolumn{3}{c}{Classical results} & \multicolumn{2}{c}{Bayesian estimates} &\multicolumn{3}{c}{Posterior probabilities}\\
\hline
                        & MLE & s.e. & $p$ & post. mean & post. sd & $P(\beta=0|\textbf{Y})$ &  $P(\beta<0|\textbf{Y})$ & $P(\beta>0|\textbf{Y})$\\
edges                    & -4.039  &    1.290  &  0.002 &-2.525& 0.675 &    - &    - &    -\\
pref. sim. (st.)         &  0.118  &    0.079  &  0.136 & 0.116& 0.106 & 0.722 & 0.037 & 0.241\\
reciprocity                   &  0.808  &    0.248  &  0.001 & 0.765& 0.298 & 0.138 & 0.005 & 0.857\\
int. group homophily        &  1.067  &    0.291  &  0.000 & 1.222& 0.485 & 0.178 & 0.005 & 0.817\\
gov. target  &  0.597  &    0.189  &  0.002 & 0.561& 0.248 & 0.259 & 0.009 & 0.732\\
sci. source  &  0.072  &    0.218  &  0.742 & 0.063& 0.269 & 0.822 & 0.072 & 0.106\\
common committees (st.)       &  0.727  &    0.122  &  0.000 & 0.839& 0.173 & 0.000 & 0.000 & 1.000\\
sci. communication           &  2.910  &    0.630  &  0.000 & 2.935& 1.006 & 0.039 & 0.002 & 0.958\\
infl. attr. (st.)          &  0.439  &    0.088  &  0.000 & 0.438& 0.114 & 0.003 & 0.000 & 0.997\\
GWESP(0.1)      &  2.552  &    1.129  &  0.024 & 1.363& 0.597 & 0.094 & 0.012 & 0.895\\
GWDSP(0.1)      & -0.134  &    0.049  &  0.007 &-0.208& 0.087 & 0.146 & 0.847 & 0.007\\
\hline
\end{tabular}}
\label{taBF_AppA}
\end{center}
\end{table}

The secondary question in this application was the test of the order-constrained hypothesis of shared committees followed by influence attribution, followed by preference similarity versus a (null) hypothesis of equal effects versus the complement that neither of these two hypotheses hold; see equation \eqref{test2_AppA}. The edge covariates were standardized to be able to compare the effects on the same unified scale. The Bayes factors between the hypotheses and the posterior probabilities (when assuming equal prior probabilities of $\frac{1}{3}$ for each hypothesis) are given in Table \ref{taBF_AppA_test2}). We see that the order-constrained hypothesis receives approximately 74 and 98 times more evidence than the equality hypothesis $H_2$ and the complement $H_3$, respectively. These results also display the largest posterior probability for the order-constrained hypothesis $H_1$ with approximately 98\%. Thus, after observing the data, there is a strong belief that the (standardized) effect of shared committees is strongest, followed by influence attribution, followed by preference similarity. These results are also in agreement with the estimates of the coefficients in Table \ref{taBF_AppA}.

\begin{table}[t]
\begin{center}
\caption{Bayes factors between the hypotheses in the second test in Application A, and posterior probabilities using equal prior probabilities.}
{\small \begin{tabular}{lccccccc}
  \hline
Hypotheses&  \multicolumn{3}{c}{Bayes factors} & posterior probabilities\\
 & $H_1$ & $H_2$ & $H_3$ & ~  $P(H_t | \textbf{Y})$\\
\hline
$H_1:\beta_{\text{committee}} > \beta_{\text{infl att}} > \beta_{\text{pref sim}}$ & 1.000 &74.335 &97.953 & .977\\
$H_2:\beta_{\text{committee}} = \beta_{\text{infl att}} = \beta_{\text{pref sim}}$ & 0.013 & 1.000&  1.318 & .013\\
$H_3: \text{complement}$ & 0.010&   0.759 & 1.000 & .010\\
\hline
\end{tabular}}
\label{taBF_AppA_test2}
\end{center}
\end{table}

\begin{figure}[t]
\begin{center}
\includegraphics[width=15.5cm]{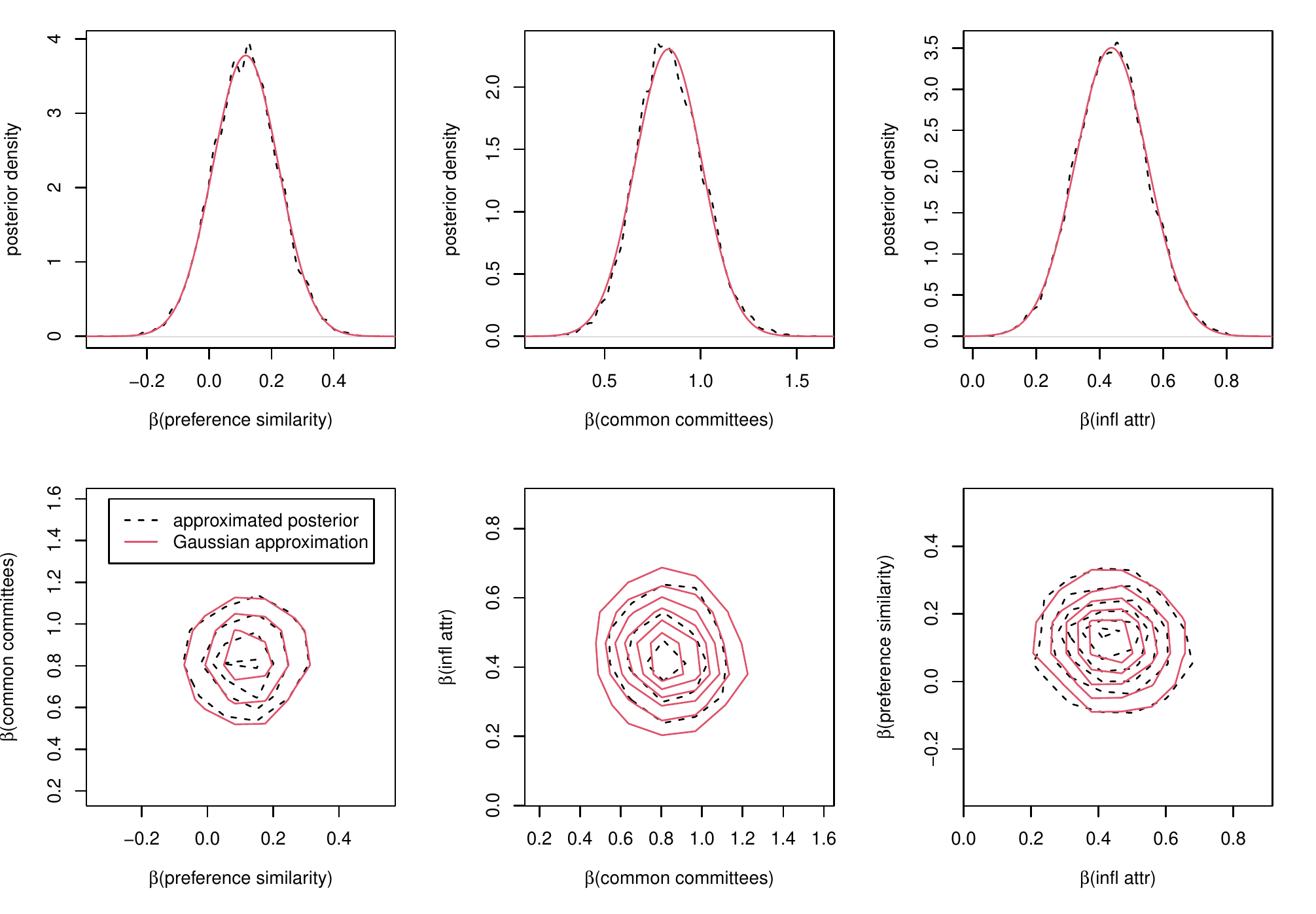}
\caption{Posterior density estimate based on 220,000 posterior draws (dashed black line) and the Gaussian approximation of the posterior (solid red line) for the policy network.}
\label{fig_AppA_approx}
\end{center}
\end{figure}

We end this application by showing the accuracy of the Gaussian approximation of the posterior of preference similarity, common committees, and influence attribution based on the posterior draws from the \texttt{bergm} function. Figure \ref{fig_AppA_approx} plots the estimated posterior (dashed black line) and the Gaussian approximation (red solid line). The figure shows that the posteriors can be well approximated using Gaussian distributions.

\subsection*{Application B: Institutional and functional overlap in co-\-author\-ship networks}
The interest is in testing competing hypotheses about an increasing positive effect of functional overlap of the form
\[
\begin{array}{l}
H_1:\beta_{\text{supervision}}>\beta_{\text{same team}}>\beta_{\text{same affiliation}}>0\\
H_2:\beta_{\text{supervision}}=\beta_{\text{same team}}=\beta_{\text{same affiliation}}>0\\
H_3:\beta_{\text{supervision}}=\beta_{\text{same team}}=\beta_{\text{same affiliation}}=0\\
H_4:\text{complement,}\\
\end{array}
\]
which are tested under the Swiss collaboration network data from \cite{leifeld2018polarization}\footnote{The \texttt{R} code for analyses can be found in Appendix \ref{AppB}.}.
The Bayes factors between the hypotheses can be found in the evidence matrix in Table \ref{taBF_evidenceB}. We see that hypothesis $H_1$, which assumes an increasing positive collegial effect, receives decisive evidence against $H_2$ (constant positive collegial effect) and $H_1$ (no collegial effect), and strong evidence against the complement hypothesis. The resulting posterior probabilities are printed in the last column of Table \ref{taBF_evidenceB} when using equal prior probabilities. We see that the posterior probabilities for $H_1$, $H_2$, $H_3$, and $H_4$ are approximately 91\%, 2\%, 0\%, and 7\%. Thus, $H_1$, which assumes an increasing positive effect of functional overlap, is clearly most likely after observing the data. Furthermore, we can safely rule out hypothesis $H_3$, which assumed no functional overlap effect, given the zero posterior probability. Again, we note that no decision needs to be made which hypothesis is true because these outcomes can be interpreted on a continuous scale. But if it was concluded that hypothesis $H_1$ is true, there would be a conditional probability of drawing the wrong conclusion of about 9\% given the observed network.

\begin{table}[t]
\begin{center}
\caption{Bayes factors between the hypotheses in Application B, and posterior probabilities based on equal prior probabilities.}
{\small
\hspace*{-2cm}
\begin{tabular}{lccccccc}
  \hline
Hypotheses &  \multicolumn{4}{c}{Bayes factors} & posterior probabilities\\
& $H_1$ & $H_2$ & $H_3$ & $H_4$ & ~  $P(H_t | \textbf{Y})$\\
\hline
$H_1:\beta_{\text{supervision}}>\beta_{\text{same team}}>\beta_{\text{same affiliation}}>0$& 1.000 & 59.686 &25612.159 &12.319 & .911\\
$H_2:\beta_{\text{supervision}}=\beta_{\text{same team}}=\beta_{\text{same affiliation}}>0$& 0.017 &   1.000 &  429.117 & 0.206 & .015\\
$H_3:\beta_{\text{supervision}}=\beta_{\text{same team}}=\beta_{\text{same affiliation}}=0$& 0.000 & 0.002    & 1.000 & 0.000 & .000\\
$H_4:\text{complement}$& 0.081 & 4.845 & 2079.047 & 1.000 & .074\\
\hline
\end{tabular}}
\label{taBF_evidenceB}
\end{center}
\end{table}

\begin{figure}[t]
\begin{center}
\includegraphics[width=16cm]{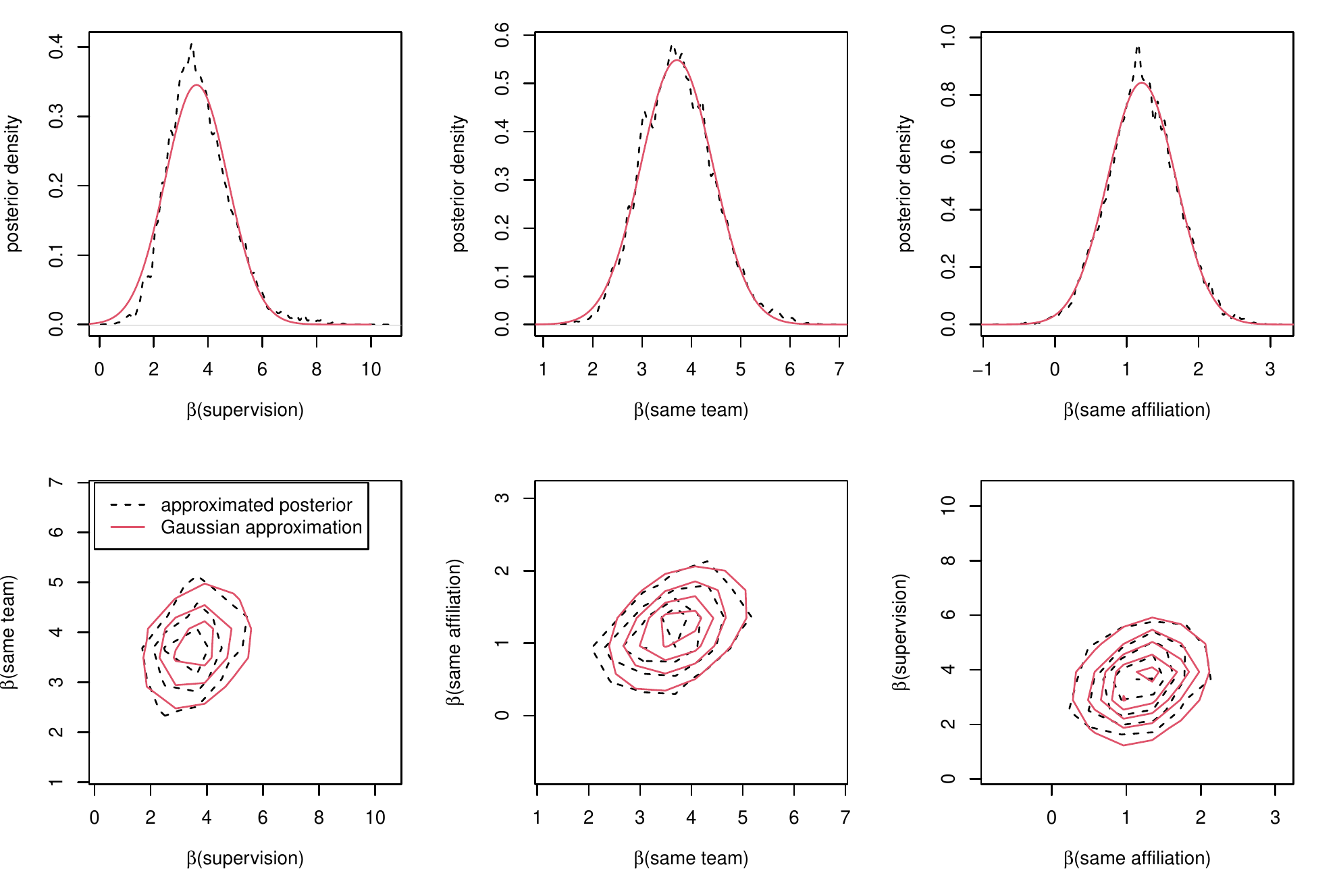}
\caption{Posterior density estimate based on 1,000,000 posterior draws (dashed black line) and the Gaussian approximation of the posterior (solid red line) for the co-authorship network.}
\label{fig_AppBF_approx}
\end{center}
\end{figure}

A visual check was performed to see how well the posterior distribution of the key parameters in this hypothesis test can be approximated with a Gaussian distribution. In Figure \ref{fig_AppBF_approx}, the black dashed lines are the estimated posteriors, and the solid red lines are the Gaussian approximations. Overall, the Gaussian approximations seem accurate with a slight deviation for the posterior of the supervision effect due to its skewness. Based on these results, we conclude that the Gaussian approximation is sufficiently accurate.

Finally, we present the classical result, Bayesian estimates, and posterior probabilities of the hypotheses for all separate coefficients in Table \ref{taBF_AppB}. We see some differences between the Bayesian and the classical results. Several coefficients result in relatively small $p$-values while the Bayesian posterior probability of a zero effect for these coefficients are substantial, such as GWDegree, gender, share English articles, number of publications, and English article similarity. These differences can be explained by the relatively large posterior standard deviations in comparison to the classical standard errors implying more posterior uncertainty regarding the true parameter value. Again, we see that some Bayesian estimates are considerably larger than the classical estimates. The table also shows that the Bayesian estimate effect of same team is slightly larger than the estimated effect of supervisor (both having large posterior standard deviations), which conflicts with the constraint $\beta_{\text{supervision}}>\beta_{\text{same team}}$ under $H_1$. The clear support in favor of $H_1$, however, can be explained by the fact that a very slight violation of one order constraint can be compensated by clear support for the other order constraints. We can also see this when looking at the posterior and prior probability that the constraints of $H_1$ hold under the full unconstrained ERGM, i.e., the numerator and denominator of the second term on the right hand side of Equation \eqref{Btu}, which are equal to 0.460 and .064, respectively\footnote{These values are printed in the \texttt{Specification table} when calling \texttt{summary()} of the \texttt{BF} object \citep[see also][]{mulder2021bfpack}.}. Thus, the relative fit is much larger than the relative complexity of $H_1$ resulting in clear support for the the order-constrained hypothesis \citep[see also][]{hoijtink2019tutorial}.

\begin{table}[t]
\begin{center}
\caption{Classical results (MLE, standard error, and $p$-value) and posterior probabilities of a null, a negative, and a positive effect (assuming equal prior probabilities) for all ERGM coefficients of the co-authorship network (Application B).}
{\small
\hspace*{-2.2cm}
\begin{tabular}{lccccccccccc}
  \hline
&  \multicolumn{3}{c}{Classical results} & \multicolumn{2}{c}{Bayesian estimates} & \multicolumn{3}{c}{Posterior probabilities}\\
\hline
                        &    MLE & s.e.  &    $p$ & post. mean & post. sd & $P(\beta=0|\textbf{Y})$ &  $P(\beta<0|\textbf{Y})$ & $P(\beta>0|\textbf{Y})$\\
edges                        & -10.099   &     0.625 &   0.000 & -9.010& 0.842&     -  &   -  &   - \\
supervision       &   2.864  &    0.367  &  0.000 & 3.692 &1.103 & 0.035 & 0.001 & 0.965 \\
same team         &   2.594  &    0.298  &  0.000 & 3.769 &0.681 & 0.000 & 0.000 & 1.000 \\
same affiliation  &   1.052  &    0.229  &  0.000 & 1.210 &0.459 & 0.109 & 0.004 & 0.887 \\
GWESP             &   1.507   &     0.164 &   0.000 & 1.552 &0.240&  0.000&  0.000 & 1.000 \\
GWDegree          &   0.983  &    0.286  &  0.001 & 0.406& 0.385 & 0.569 & 0.061 & 0.370 \\
seniority         &   0.125  &    0.120  &  0.297 & 0.288& 0.271 & 0.630 & 0.054 & 0.316 \\
gender            &   0.354  &    0.184  &  0.053 &-0.031& 0.297 & 0.727 & 0.151 & 0.122 \\
gender homophily  &   0.042  &    0.234  &  0.858 & 0.192& 0.393 & 0.710 & 0.092 & 0.197 \\
geographic distance  &  -0.239  &    0.150  &  0.111 &-0.502& 0.288 & 0.350 & 0.622 & 0.028 \\
topic similarity  &  15.476  &    2.120  &  0.000 &16.782& 4.590 & 0.005 & 0.000 & 0.994 \\
share Eng. articles    &   0.407 &     0.184  &  0.027 & 0.263& 0.419 & 0.708 & 0.077 & 0.215 \\
number of publications       &   0.022  &    0.005  &  0.000 & 0.019& 0.011 & 0.428 & 0.024 & 0.547 \\
Eng. art. similarity    &   2.038   &   0.495  &  0.000 & 1.432& 0.857 & 0.350 & 0.032 & 0.618 \\
\hline
\end{tabular}}
\label{taBF_AppB}
\end{center}
\end{table}

\section{Discussion}
A Bayes factor was proposed for testing a broad class of statistical hypotheses with competing equality and/or order constraints on the coefficients under ERGMs. The outcome of the Bayes factor quantifies the relative evidence in the data in favor of a hypothesis with a set of constraints on the ERGM coefficients relative to a hypothesis with alternative constraints on the coefficients. When prior probabilities for the hypotheses are formulated, the Bayes factors can be translated to posterior probabilities of the hypotheses, which quantify the plausibility of the hypothesis after observing the data. Several attractive properties of the methodology were illustrated, such as its consistent behavior, its ability to test multiple hypotheses simultaneously, its ability to quantify the evidence in the data in favor of the null, and its ability to balance between fit and complexity as an Occam's razor, also for hypotheses with order constraints. These properties are not shared by classical $p$-values (which are currently dominant for hypothesis testing in applied social network research using ERGMs). Consequently, when observing a nonsignificant testing outcome, this does not tell us whether this implies there is evidence of absence of an effect (i.e., support for the simpler model) or absence of evidence (i.e., inconclusive evidence). Bayes factors, on the other hand, are able to distinguish between these two scenarios.

The proposed Bayes factors were based on a $g$-prior that contains the information of one dyad \citep{Zellner:1986,Liang:2008}. A flat prior was specified for the edges parameter (which serves as an intercept) together with a multivariate normal prior for the ERGM coefficients having zero means and a prior covariance structure that was based on the design matrix of the change statistics of the full ERGM. Zero prior means imply that the prior is neutral regarding the direction of the effect. Thereby the prior differs from the unit-information on which the BIC is based which is centered at the MLE \citep{Schwarz:1978,Raftery:1995}. When testing simple null hypotheses of whether an effect equals zero, the proposed Bayes factor was less conservative towards the null in comparison to the BIC \citep[which is sometimes a critique of the BIC, e.g., see][]{weakliem1999critique}. The AIC, on the other hand, was more liberal than the proposed Bayes factor, which can be explained by the fact that the AIC is not consistent as a measure of evidence if a null model is true \citep{OHagan:1995}.

To test equality/order-constrained hypotheses, truncated versions of the full unconstrained prior were constructed, so that the Bayes factors between any pair of hypotheses can be computed extremely fast using extended Savage-Dickey ratios and Gaussian approximations of the (kernel of the) posterior. This prior approach has also been considered by \cite{Klugkist:2007} for testing constrained hypotheses, for instance. Furthermore, these truncated priors naturally appear in partial Bayes factors and fractional Bayes factors \citep{Berger:1996,OHagan:1995,Mulder:2010,Mulder:2014b,mulder2022bayesian}. For interesting discussions of this approach and possible alternatives (in the realm of linear models), we refer the interested reader to \cite{consonni2008compatibility}, \cite{Wetzels:2010}, \cite{marin2010resolving}, \cite{heck2019caveat}, \cite{mulder2021prevalence}, \cite{mulder2022generalization}, and the references therein.

Furthermore, the two empirical applications showed some differences between the classical estimates and the Bayesian estimates based on the proposed unit-information prior. Even though one might expect that the Bayesian estimates are closer to zero (due to prior shrinkage towards the prior mean of zero), some Bayesian estimates and posterior standard deviations were larger than their classical counterparts. We also observed this when fitting a Bayesian ERGM using flat priors. Therefore the observed differences between the Bayesian and classical estimates are not necessarily attributed to the proposed prior. In summary, prior specification under ERGMs remains an important topic for Bayesian hypothesis testing and Bayesian estimation problems, which deserves further attention in the literature.

Finally, the implementation of the new methodology in the \texttt{R} package \texttt{BFpack} \citep{mulder2021bfpack} allows researchers to apply the methodology in a straightforward manner. Thereby, the paper contributes to the increasing availability of Bayes factors for social network analysis, such as network autocorrelation models \citep{dittrich2019network,dittrich2020network}, relational event models \citep{mulder2019modeling}, as well as other statistical frameworks such as structural equation models \citep{Gu:2019,van2021teacher}, or (multivariate) regression models \citep{rouder2012default,braeken2015relative,mulder2022bayesian}. For a recent overview of applications of Bayes factor testing in the social and behavioral sciences, we refer interested readers to \cite{heck2022review}. Furthermore, tutorials on the use of Bayes factors are available by \cite{Masson:2011}, \cite{wagenmakers2018bayesian}, or \cite{hoijtink2019tutorial}.

\section*{Acknowledgements} The authors wish to thank two anonymous reviewers for their constructive feedback which helped to improve the manuscript. The first author was supported by an ERC Starting Grant (758791).

\bibliographystyle{apacite}
\bibliography{refs_mulder}

\appendix

\section{R code for the analysis of Application A}\label{AppA}
\begin{verbatim}
install.packages("BFpack")

library("statnet")
library("ergm")
library("BFpack")
library("btergm")

seed <- 1234

set.seed(seed)

data("chemnet") # see ?chemnet more details on the data 
                # (Leifeld & Schneider, 2012, AJPS)

# create confirmed network relation
sci <- scito * t(scifrom)  # equation 1 in the AJPS paper
prefsim <- dist(intpos, method = "euclidean")  # equation 2
prefsim <- max(prefsim) - prefsim  # equation 3
prefsim <- as.matrix(prefsim)
prefsim_scaled <- c(scale(c(prefsim[lower.tri(prefsim)],
                prefsim[upper.tri(prefsim)])))
# standardize certain predictors (for the order test on the ergm coefficients)
prefsim_st <- prefsim
prefsim_st[lower.tri(prefsim)] <- prefsim_scaled[1:(length(prefsim_scaled)/2)]
prefsim_st[upper.tri(prefsim)] <- prefsim_scaled[(length(prefsim_scaled)/2+1):
                length(prefsim_scaled)]
committee <- committee %*% t(committee)  # equation 4
diag(committee) <- 0 # the diagonal has no meaning
committee_scaled <- c(scale(c(committee[lower.tri(committee)],
                committee[upper.tri(committee)])))
committee_st <- committee
committee_st[lower.tri(committee)] <- committee_scaled[1:
                (length(committee_scaled)/2)]
committee_st[upper.tri(committee)] <- committee_scaled[
                (length(committee_scaled)/2+1):length(committee_scaled)]
types <- types[, 1]  # convert to vector
infrep_st <- matrix(c(scale(c(infrep))),nrow=nrow(infrep))

# create network objects and store attributes
nw.pol <- network(pol) # political/stratgic information exchange
set.vertex.attribute(nw.pol, "orgtype", types)
set.vertex.attribute(nw.pol, "betweenness",
                     betweenness(nw.pol)) # centrality

nw.sci <- network(sci) # technical/scientific information exchange
set.vertex.attribute(nw.sci, "orgtype", types)
set.vertex.attribute(nw.sci, "betweenness",
                     betweenness(nw.sci)) # centrality

# Model 2 of the AJPS paper
model2 <- ergm(nw.pol ~ edges +
                 edgecov(prefsim_st) +
                 mutual +
                 nodemix("orgtype", base = -7) +
                 nodeifactor("orgtype", base = -1) +
                 nodeofactor("orgtype", base = -5) +
                 edgecov(committee_st) +
                 edgecov(nw.sci) +
                 edgecov(infrep_st) +
                 gwesp(0.1, fixed = TRUE) +
                 gwdsp(0.1, fixed = TRUE),
               control = control.ergm(seed = seed)
)
summary(model2)

# see names of the coefficients on which constrained hypotheses can be 
# formulated using BF() function
get_estimates(model2)
# Test hypothesis
BF_AppA <- BFpack::BF(model2,hypothesis="edgecov.prefsim = 0",main.iters = 10000)
summary(BF_AppA)

#get Bayesian estimates using the unit information prior
BF_AppA$estimates

# To compute the evidence based on the BIC and AIC, we also need to fit the null
# model (H1 in the paper) without preference similarity
model2_H0 <- ergm(nw.pol ~
                    edges +
                    mutual +
                    nodemix("orgtype", base = -7) +
                    nodeifactor("orgtype", base = -1) +
                    nodeofactor("orgtype", base = -5) +
                    edgecov(committee_st) +
                    edgecov(nw.sci) +
                    edgecov(infrep_st) +
                    gwesp(0.1, fixed = TRUE) +
                    gwdsp(0.1, fixed = TRUE),
                    control = control.ergm(seed = seed)
)
#BF01-BIC
exp(-BIC(model2_H0)[1]/2 + BIC(model2)[1]/2)
#BF01-paper
BF_AppA$BFtu_exploratory[1,1]
#ratio01-AIC
exp(-AIC(model2_H0)[1]/2 + AIC(model2)[1]/2)

# second order test in Application A
hypA <- "edgecov.committee_st > edgecov.infrep_st > edgecov.prefsim_st;
         edgecov.committee_st = edgecov.infrep_st = edgecov.prefsim_st"
BF_AppA_test2 <- BFpack::BF(model2,
                    hypothesis=hypA,
                    main.iters = 10000)
summary(BF_AppA_test2)


\end{verbatim}

\section{R code for the analysis of Application B}\label{AppB}
\begin{verbatim}
install.packages("BFpack")

library("network")
library("ergm")
library("BFpack")
library("btergm")

seed <- 1234

data(ch_coauthor) # see ?ch_coauthor more details on the data 
                  # (Leifeld, 20128, Physics A)

# ERGM for Switzerland, full model
ch_nw <- network(ch_coaut, directed = FALSE)
set.vertex.attribute(ch_nw, "frequency", ch_nodeattr$num_publications)
set.vertex.attribute(ch_nw, "status", as.character(ch_nodeattr$status))
set.vertex.attribute(ch_nw, "male", ch_nodeattr$male)
set.vertex.attribute(ch_nw, "share_en_articles", ch_nodeattr$share_en_articles)

ch_inst_indices <- which(grepl("^inst_.+", colnames(ch_nodeattr)))
ch_same_affiliation <- as.matrix(ch_nodeattr[, ch_inst_indices]) %*% 
                  t(ch_nodeattr[, ch_inst_indices])

ch_nodeattr$chairtitle[ch_nodeattr$chairtitle == ""] <- NA
ch_same_chair <- matrix(0, nrow = nrow(ch_same_affiliation), ncol = 
                  ncol(ch_same_affiliation))  # matrix: same chair
for (i in 1:length(ch_nodeattr$chairtitle)) {
  for (j in 1:length(ch_nodeattr$chairtitle)) {
    if (i != j && !is.na(ch_nodeattr$chairtitle[i]) && !is.na(
                  ch_nodeattr$chairtitle[j]) && ch_nodeattr$chairtitle[i] 
                  == ch_nodeattr$chairtitle[j]
        && ch_same_affiliation[i, j] == TRUE) {
      ch_same_chair[i, j] <- 1
    }
  }
}
rownames(ch_same_chair) <- rownames(ch_same_affiliation)
colnames(ch_same_chair) <- colnames(ch_same_affiliation)

ch_supervision <- ch_same_affiliation *
  ch_same_chair *
  matrix(ch_nodeattr$status == "professor",
         nrow = nrow(ch_same_chair),
         ncol = ncol(ch_same_chair),
         byrow = FALSE) *
  matrix(ch_nodeattr$status != "professor",
         nrow = nrow(ch_same_chair),
         ncol = ncol(ch_same_chair),
         byrow = TRUE)

# parameterization in terms of groups with dummy's
ch_same_affiliation1 <- ch_same_affiliation - ch_same_chair
ch_same_chair1 <- ch_same_chair - ch_supervision
ch_model_para <- ergm(ch_nw ~
                         edges +
                         gwesp(0.3, fixed = TRUE) +
                         gwdegree(0.4, fixed = TRUE) +
                         nodecov("frequency") +
                         nodefactor("status") +
                         nodefactor("male") +
                         nodematch("male") +
                         edgecov(ch_dist100km) +
                         edgecov(ch_same_affiliation1) +
                         edgecov(ch_same_chair1) +
                         edgecov(ch_supervision) +
                         edgecov(ch_topicsim) +
                         nodecov("share_en_articles") +
                         edgecov(ch_en_article_sim),
                         control = control.ergm(seed = seed))
summary(ch_model_para)
# hypothesis formulation
hypB <-"0<edgecov.ch_same_affiliation1<edgecov.ch_same_chair1<edgecov.ch_supervision;
    0<edgecov.ch_same_affiliation1=edgecov.ch_same_chair1=edgecov.ch_supervision;
    edgecov.ch_same_affiliation1=edgecov.ch_same_chair1=edgecov.ch_supervision=0"
# below the 'main.iters' argument is used to get more posterior draws and thus
# a numerically more accurate outcome. By omitting it the default of 1000 draws
# is used which is faster (but numerically less accurate).
BF_AppB1_para <- BFpack::BF(ch_model_para,
                         hypothesis=hypB,
                         main.iters = 50000)
summary(BF_AppB1_para)

#get Bayesian estimates using the unit information prior
BF_AppB1_para$estimates

\end{verbatim}

\end{document}